\documentclass[prd,nofootinbib,english,a4paper]{revtex4-2}
\usepackage{amsmath}
\usepackage{amsfonts}
\usepackage{amssymb}
\usepackage{xcolor}
\usepackage{float}
\usepackage{accents}
\usepackage{hyperref}
\usepackage{bookmark}
\usepackage{tabularray}
\usepackage[utf8]{inputenc}
\usepackage{graphicx}
\hypersetup{
	colorlinks=true,
	linkcolor=blue,
	filecolor=magenta,
	citecolor=blue
}
\allowdisplaybreaks

\DeclareMathOperator{\arcsinh}{arcsinh}
\DeclareMathOperator{\arccosh}{arccosh}
\DeclareMathOperator{\arccot}{arccot}
\DeclareMathOperator{\arctanh}{arctanh}
\DeclareMathOperator{\sech}{sech}

\newcommand{\dd}{\mathrm{d}}

\newcommand{\lc}[1]{\accentset{\circ}{#1}}

\begin{document}

	\title{Spherically symmetric vacuum solutions in 1-Parameter New General Relativity and their phenomenology}

	\author{Helen Asuküla}
	\email{helen.asukula@gmail.com}
	\affiliation{Laboratory of Theoretical Physics, Institute of Physics, University of Tartu, W. Ostwaldi 1, 50411 Tartu, Estonia}

	\author{Sebastian Bahamonde}
	\email{sbahamondebeltran@gmail.com,sebastian.bahamonde@ipmu.jp}
	\affiliation{Kavli Institute for the Physics and Mathematics of the Universe (WPI), The University of Tokyo Institutes
		for Advanced Study (UTIAS), The University of Tokyo, Kashiwa, Chiba 277-8583, Japan.}

	\author{Manuel Hohmann}
	\email{manuel.hohmann@ut.ee}
	\affiliation{Laboratory of Theoretical Physics, Institute of Physics, University of Tartu, W. Ostwaldi 1, 50411 Tartu, Estonia}

	\author{Vasiliki Karanasou}
	\email{vasiliki.karanasou@ut.ee}
	\affiliation{Laboratory of Theoretical Physics, Institute of Physics, University of Tartu, W. Ostwaldi 1, 50411 Tartu, Estonia}

	\author{Christian Pfeifer}
	\email{christian.pfeifer@zarm.uni-bremen.de}
	\affiliation{ZARM, University of Bremen, 28359 Bremen, Germany}

	\author{Jo\~{a}o Lu\'{i}s Rosa}
	\email{joaoluis92@gmail.com}
	\affiliation{Institute of Physics, University of Tartu, W. Ostwaldi 1, 50411 Tartu, Estonia}
	\affiliation{Institute of Theoretical Physics and Astrophysics, University of Gda\'{n}sk, Jana Ba\.{z}y\'{n}skiego 8, 80-309 Gda\'{n}sk, Poland}

	\begin{abstract}
	In this work, we study spherically symmetric vacuum solutions in 1-parameter New General Relativity (NGR), a specific theory in teleparallel gravity which is constructed from the three possible quadratic scalars obtained from torsion with arbitrary coefficients satisfying the requirements for the absence of ghosts. In this class of modified theories of gravity, the observable effects of gravity result from the torsion rather than the curvature of the spacetime. Unlike in GR, where the fundamental quantity is the metric from which the Levi-Civita connection is derived, in teleparallel theories of gravity the fundamental variable is the tetrad, from which one constructs the metric and the teleparallel connection. We consider the most general tetrad for spherical symmetry and we derive the corresponding field equations. Under adequate assumptions, we find three different branches of vacuum solutions and discuss their associated phenomenology. In particular, we analyze the photon sphere, the classical tests of GR such as the light deflection, the Shapiro delay, and the perihelion shift, and also the Komar mass, while providing a detailed comparison with their Schwarzschild spacetime counterparts. Finally, we analyze how the observational imprints from accretion disks and shadows are affected in comparison with their GR counterparts, and conclude that the free parameters of the model might induce additional attractive or repulsive effects to the propagation of photons, depending on their values.
	\end{abstract}

	\maketitle



	\section{Introduction}
	Since its publication in 1915, the theory of General Relativity (GR) has been proven successful in accounting for a wide variety of phenomena from the astrophysical and cosmological point of view \cite{Will:2014kxa,Will:2018bme,Baker:2014zba} from the classical tests of GR \cite{Einstein:1916vd} to the modern post-Newtonian tests, as well as as experiments involving the time delay of light, gravitational lensing, and the equivalence principle \cite{1959Sci...129..621D,1960AmJPh..28..340S,PhysRevD.66.082001,2009ApJ...699.1395F,1964PhRvL..13..789S,PhysRev.170.1186,PhysRevLett.93.261101}. Nevertheless, the recent discovery of the late-time cosmic acceleration \cite{SupernovaCosmologyProject:1998vns,SupernovaSearchTeam:1998fmf} has risen important theoretical problems in modern cosmology. Although the standard $\Lambda$CDM model of cosmology, where dark energy is described by a cosmological constant $\Lambda$ alongside cold dark matter (CDM) is favored by observations~\cite{Copeland:2006wr,Planck:2018vyg}, it also exhibits observational tensions~\cite{DiValentino:2021izs,Schoneberg:2021qvd,Abdalla:2022yfr}, which motivate the study of an alternative scenarios which propose that this cosmological behavior arises as a consequence of modifications to the gravitational theory itself~\cite{Clifton:2011jh,Nojiri:2010wj,Heisenberg:2018vsk,CANTATA:2021ktz}.

	There are several ways to modify gravity. In GR, the effects of gravity are the result of the curvature of the spacetime, which depends on the Levi-Civita connection. A simple generalization of GR, the $f(R)$ theory of gravity, is obtained by generalizing the linear dependency of the gravitational action on the Ricci scalar $R$ to an arbitrary function $f\left(R\right)$ \cite{Capozziello:2011et,Sotiriou:2008rp}. If we assume a general affine connection, the quantities torsion and non-metricity are also introduced along with the curvature~\cite{Hehl:1994ue,JimenezCano:2021rlu}. A particularly interesting class of these theories are teleparallel theories of gravity, where the gravitational field is mediated by either the torsion or non-metricity~\cite{BeltranJimenez:2019esp}, or both~\cite{BeltranJimenez:2019odq}, while the curvature vanishes. In the following, we focus on metric teleparallel theories, where the non-metricity vanishes, and only torsion is present. While in GR the fundamental variable is the metric, from which the Levi-Civita connection is derived, the fundamental variables in a teleparallel theory are the tetrad and the spin connection, which in turn define the metric and the teleparallel affine connection~\cite{Aldrovandi:2013wha,Bahamonde:2021gfp}. It is always possible though to work on the Weitzenböck gauge where the spin connection vanishes \cite{Krssak:2018ywd}.

	The simplest teleparallel theory of gravity is the so-called Teleparallel Equivalent of GR (TEGR) for which the corresponding Lagrangian depends on the torsion scalar $T$. A commonly considered and quite simple generalization of the TEGR is the New General Relativity (NGR) theory \cite{Hayashi:1979qx}. In this theory, the torsion scalar is decomposed into its irreducible components, and three free parameters are introduced in the Lagrangian. Different values of these parameters correspond to different gravitational theories. In fact, it has been proved that in order to reproduce the Newtonian limit, two of the parameters should have fixed values, the ones from TEGR, and we are left with a single free parameter. This corresponds to the 1-parameter New General Relativity (1PNGR) \cite{amendum}. Several other  modifications have been studied, e.g. $f(T)$ is the analogue of $f(R)$ in the teleparallel context. However, despite the fact that TEGR is equivalent to GR, its generalization $f(T)$ is not equivalent to $f(R)$ gravity. Other examples include the $f(T,B)$ gravity, where $B$ is a boundary term between the Ricci scalar and the torsion scalar, scalar-torsion theories and the Gauss-Bonnet teleparallel theories (see \cite{Bahamonde:2021gfp} for a review).

	Several topics in NGR have been already investigated, in particular the strong coupling problem and instabilities around Minkowski spacetime \cite{BeltranJimenez:2019nns}, the Hamiltonian formalism \cite{Mitric:2019rop,Cheng:1988zg, Okolow:2011np, Blixt:2018znp,Hohmann:2019sys, Blixt:2019mkt, Blixt:2019ene,Guzman:2020kgh, Blixt:2020ekl}, the Parameterized post-Newtonian limit \cite{Ualikhanova:2019ygl}, the polarization of gravitational waves \cite{Hohmann:2018xnb} as well as their propagation \cite{Hohmann:2018jso} and a field theory approach \cite{Mylova:2022ljr}. In order to gain a better understanding of the gravitational theory we are interested in and its viability and to obtain more information about black holes and other compact objects, it is useful to study the spherical solutions of the theory. Spherical solutions have been investigated in various teleparallel extensions such as $f(T,B)$ gravity \cite{Bahamonde:2021srr}, scalar-torsion gravity \cite{Bahamonde:2022lvh} and teleparallel scalar Gauss-Bonnet gravity \cite{Bahamonde:2022chq}. Solutions with spherical symmetry have also been found in 3-parameter NGR \cite{Golovnev:2023uqb}. However, the tetrad that was used is not the most general one and thus, only one branch of solutions was found.

	In this work, we use the most general tetrad ~\cite{Hohmann:2019nat,Hohmann:2019fvf}, we derive the spherically symmetric field equations and we find three branches of vacuum solutions which we investigate in 1PNGR. The first branch is the one found and analyzed in the aforementioned paper and this solution leads to the Schwarzschild metric. The second branch corresponds to a teleparallel modification of the Schwarzschild metric. Here, the $rr$-component of the metric differs from the inverse of the $tt$-component  by a constant. At first sight this might look like a mild modification. However, it leads to the fact that this spacetime is no longer asymptotically flat. We also discuss phenomenological aspects of the solution found for the second branch which has two free parameters. In particular, we investigate the existence of horizons and singularities and we compute the Komar mass. Furthermore, the physical relevance of these solutions in an astrophysical context is assessed through an analysis of a few classical tests of GR \cite{Einstein:1916vd}, namely the phenomena of light deflection, Shapiro time delay and perihelion shift, as well as the more recent experimental observation of the black hole (BH) shadow by the EHT collaboration \cite{EventHorizonTelescope:2019dse,EventHorizonTelescope:2020qrl,EventHorizonTelescope:2022wkp}, the latter analyzed recurring to a numerical ray-tracing code widely used in the literature \cite{Rosa:2023qcv,Rosa:2023hfm,Olmo:2023lil,Rosa:2022tfv,daSilva:2023jxa,Guerrero:2022msp,Guerrero:2022qkh,Olmo:2021piq,Guerrero:2021ues}.

	The paper is organized in the following way: In Section \ref{sec:tele_intro}, the fundamentals of teleparallel gravity and 1PNGR are introduced. In Section \ref{sec:sph} we introduce spherical symmetry. We solve the antisymmetric field equations immediately  in Section \ref{sec:asym}, where we find three branches of solutions, and derive the symmetric field equations for each branch in Section \ref{sec:sym}. In Section \ref{sec:sph_sol} the spherical vacuum solutions are found for the three branches. In Section \ref{sec:Obs}, the observables in spherical symmetry are studied. We summarize the conclusions of this work in Section \ref{sec:Conclusion}.

	The notational conventions in this article are as follows: Indices $A,B,...$ label frames/tetrads and $\mu, \nu, ...$ label coordinates; both run from $0$ to $3$. The Minkowski metric components are denoted by $\eta_{AB} = \textrm{diag}(-1,1,1,1)$. A ring above quantities like $\mathring{R}$ marks objects constructed with the Levi-Civita connection of the metric.

	\section{Teleparallel gravity and 1-parameter NGR}\label{sec:tele_intro}
	In this section we briefly introduce the main concepts of teleparallel theories of gravity, 1-parameter NGR, before we discuss spherical symmetry and the field equations in spherical symmetry in Section \ref{sec:sph}.

	Teleparallel theories of gravity are formulated in terms of a tetrad $e^A = e^A{}_\mu \dd x^\mu$ and spin-connection coefficients $\omega^A{}_B = \omega^A{}_{B\mu}\dd x^{\mu}$ of a flat, metric-compatible connection with torsion components $T^A{}_{\mu\nu}$ \cite{Bahamonde:2021gfp,Pereira:2013qza,Krssak:2018ywd,Aldrovandi:2013wha}. The tetrad itself defines the metric and the components of the torsion tensor through the relations
	\begin{align}
		g_{\mu\nu} = \eta_{AB} e^A{}_\mu e^B{}_\nu,\quad
		T^A{}_{\mu\nu} = \frac{1}{2}\left( \partial_\mu e^A{}_\nu  - \partial_\nu e^A{}_\mu + \omega^A{}_{B\mu}e^B{}_\nu - \omega^A{}_{B\nu}e^B{}_\mu\right)\,.
	\end{align}
	The spin connection coefficients ensure the invariance under a simultaneous Lorentz transformation of the tetrads $e^A \to \tilde e^A = \Lambda^A{}_B e^B$ and spin connection $\omega^A{}_B\to\tilde{\omega}^A{}_{B\mu} = \Lambda^A{}_C(\Lambda^{-1})^D{}_B\,\omega^C{}_{D\mu} + \Lambda^A{}_C\,\partial_{\mu}(\Lambda^{-1})^C{}_B$. Since the connection is flat and metric compatible, it is always possible to choose the so-called Weitzenb\"ock gauge, i.e., a Lorentz frame satisfying $\omega^A{}_{B\mu} = 0$, without loss of generality. Thus, in the following we take the Weitzenb\"ock gauge and consider the tetrad as the only fundamental dynamical variable.

	In order to construct teleparallel theories of gravity, it is convenient to introduce the following quantities, defined in terms of the torsion:
	\begin{align}\label{eq:tortensatv123}
		a_{\mu} := \frac{1}{6}\epsilon_{\mu\nu\sigma\rho}T^{\nu\sigma\rho}\,,\,\quad
		v_{\mu} := T^{\sigma}{}_{\sigma\mu}\,,\,\quad
		t_{\sigma\mu\nu} := \frac{1}{2}\left(T_{\sigma\mu\nu} + T_{\mu\sigma\nu}\right) + \frac{1}{6}\left(g_{\nu\sigma}v_{\mu} + g_{\nu\mu}v_{\sigma}\right) - \frac{1}{3}g_{\sigma\mu}v_{\nu}\,,
	\end{align}
	called the axial, vector and tensor torsion, respectively. The tensor torsion satisfies the symmetries
	\begin{equation}
		t_{\alpha\mu\nu} = t_{\mu\alpha\nu}, \quad
		t_{\alpha\mu\nu} + t_{\nu\alpha\mu} + t_{\mu\nu\alpha} = 0, \quad
		t^{\alpha\mu}{}_{\alpha} = t_{\alpha}{}^{\alpha\mu} = t^{\mu\alpha}{}_{\alpha} = 0\,.
	\end{equation}
Then, the torsion tensor can be written as
\begin{equation} \label{chp4_tor_ten_decom}
	T_{\mu\nu\rho} =\frac{1}{3}(g_{\mu\nu}v_{\rho} - g_{\mu\rho}v_{\nu})+\epsilon_{\mu\nu\rho\sigma}a^{\sigma}+\frac{2}{3}(t_{\mu\nu\rho} - t_{\mu\rho\nu})\,.
\end{equation}
	From these definitions, one can construct three independent parity-even scalars
	\begin{subequations}\label{eq:torscalatv}
		\begin{align}
			T_{\text{axi}} &:= a_{\mu}a^{\mu} = \frac{1}{18}\left(2T_{\sigma\mu\nu}T^{\mu\sigma\nu} - T_{\sigma\mu\nu}T^{\sigma\mu\nu}\right)\,,\\[0.5ex]
			T_{\text{vec}} &:= v_{\mu}v^{\mu} = T^{\sigma}{}_{\sigma\mu}T_{\rho}{}^{\rho\mu}\,,\\[0.5ex]
			T_{\text{ten}} &:= t_{\sigma\mu\nu}t^{\sigma\mu\nu} = \frac{1}{2}\left(T_{\sigma\mu\nu}T^{\sigma\mu\nu} + T_{\sigma\mu\nu}T^{\mu\sigma\nu}\right) - \frac{1}{2}T^{\sigma}{}_{\sigma\mu}T_{\rho}{}^{\rho\mu}\,,
		\end{align}
	\end{subequations}
	which serve as building blocks for the Lagrangian densities that define the theories we are interested in. One particular combination of these torsion scalars is the TEGR torsion scalar given by
	\begin{equation}\label{eq:torscalar_decomposition}
		T =  \frac{3}{2}T_{\text{axi}} + \frac{2}{3}T_{\text{ten}} - \frac{2}{3}T_{\text{vec}}\,,
	\end{equation}
	which defines the teleparallel formulation of GR through the action
	\begin{equation}\label{action00}
		S_{\rm TEGR}[e^A{}_\mu]=\frac{1}{2\kappa^2}\int \dd^4x e(-T+L_{\rm m})\,,
	\end{equation}
where $e=\textrm{det}(e^A{}_\mu)$.	A variation of this action w.r.t. the tetrad components yields the field equations of TEGR, which are equivalent to the Einstein field equations in GR. 
In \cite{Hayashi:1979qx} an extension of GR called New General Relativity (NGR), obtained by replacing the numeric coefficients in Eq. \eqref{eq:torscalar_decomposition} by arbitrary parameters $c_{\rm axi}$, $c_{\rm vec}$ and~$c_{\rm ten}$, was proposed. The action describing such a theory is given by
	\begin{equation}\label{action0}
		S_{\rm NGR}[e^A{}_\mu]=\frac{1}{2\kappa^2}\int \dd^4x e (-c_{\rm vec} T_{\rm vec}-c_{\rm axi} T_{\rm ax}-c_{\rm ten} T_{\rm ten}+L_{\rm m})\,.
	\end{equation}
	It is thus clear that when the parameters $c_{\text{axi}}$, $c_{\text{ten}}$, and $c_{\text{vec}}$ take the following values,
	\begin{equation}
		c_{\text{vec}} = -\frac{2}{3}\,, \quad
		c_{\text{ten}} = \frac{2}{3}\,, \quad
		c_{\text{axi}} = \frac{3}{2}\,,
		\end{equation}
	then the combination in the action in Eq.~\eqref{action0} reduces to the scalar torsion (see Eq.~\eqref{eq:torscalar_decomposition}), and the theory is equivalent to TEGR (and, consequently, to GR). This general 3-parameter NGR theory of gravity has been investigated in some detail, and numerous constraints have been obtained from its post-Newtonian analysis \cite{Ualikhanova:2019ygl} and from the analysis of its mathematical self consistency. An important conclusion of this analysis is that, in order to prevent the appearance of ghostly modes, the coefficients $c_{\rm ten}$ and $c_{\rm vec}$ must satisfy the constraint \cite{VANNIEUWENHUIZEN1973478}
	\begin{equation}
		c_{\rm ten}+c_{\rm vec}=0\,.
	\end{equation}
	This result implies that one can, without loss of generality, parametrize the theory by the following choice of parameters
	\begin{equation}
		c_{\text{vec}} = -\frac{2}{3}\,, \quad
		c_{\text{ten}} = \frac{2}{3}\,, \quad
		c_{\text{axi}} = \frac{3}{2} + \epsilon\,,
		\end{equation}
	which leads to a Lagrangian of the form of the TEGR torsion scalar plus a correction arising from the axial part, i.e.,
	\begin{equation}
		S_{\rm 1NGR}[e^A{}_\mu]=\frac{1}{2\kappa^2}\int \dd^4x e(-T-\epsilon \,T_{\rm axi}+L_{\rm m})\,.
	\end{equation}
	The theory described by such an action is called 1-parameter NGR. This theory features the same Post-Newtonian parameters as GR (for any value of $\epsilon$) and, moreover, the theory predicts two polarization modes as in GR \cite{Hohmann:2018jso}. By taking variations with respect to the tetrad, the field equations can be cast in the following form
	\begin{equation}\label{fieldeqs}
	\kappa^2\Theta_{\mu\nu} = \lc{G}_{\mu\nu} + \epsilon\left(\frac{1}{2}a^{\rho}a_{(\rho}g_{\mu\nu)} - \frac{4}{9}\varepsilon_{\nu\alpha\beta\gamma}a^{\alpha}t_{\mu}{}^{\beta\gamma} - \frac{2}{9}\varepsilon_{\mu\nu\rho\sigma}a^{\rho}v^{\sigma} - \frac{1}{3}\varepsilon_{\mu\nu\rho\sigma}\lc{\nabla}^{\rho}a^{\sigma}\right) := \kappa^2 E_{\mu\nu}\,,
	\end{equation}
	where $\mathring{G}_{\mu}$ is the usual Einstein tensor constructed with the Levi-Civita connection of the metric, and we have defined the energy-momentum tensor as
	\begin{equation}
		\Theta_{\mu\nu}=\frac{-2}{\sqrt{-g}}\frac{\delta L_{\rm m}}{\delta g^{\mu\nu}}=e^{A}{}_{\mu}\left(\frac{1}{e}\frac{\delta L_{\rm m}}{\delta e^{A}{}_{\nu'}}\right)g_{\nu\nu'}=e^{A}{}_{\mu}\Theta_{A}{}^{\nu'}g_{\nu\nu'}\,.
	\end{equation}

	In section \ref{sec:sph_sol} we discuss the spherically symmetric vacuum solutions. To do so, we now introduce spherically symmetric teleparallel geometries.

	\section{1PNGR in spherical symmetry}\label{sec:sph}
	We now apply the 1PNGR field equations displayed in the previous section to the case of spherical symmetry. We first recall the most general spherically symmetric teleparallel geometry in section~\ref{ssec:sphertetrad}, where we also introduce a convenient parametrization. This is then used to derive the antisymmetric part of the field equations in section~\ref{sec:asym}. Assuming that these are solved, we derive the symmetric part of the field equations in section~\ref{sec:sym}.

	\subsection{Spherically symmetric tetrad ansatz}\label{ssec:sphertetrad}
	In teleparallel gravity, it is usually assumed that the tetrad and spin connection follow the same symmetries. This condition can be achieved by introducing the vector fields $Z_\zeta = Z_\zeta^\mu(x) \partial_\mu$ on the spacetime, such that the tetrad and spin connection are invariant under the flow of the fields, see \cite{Hohmann:2019nat,Hohmann:2019fvf}, yielding
	\begin{align}\label{eq:tpsymm}
		\mathcal{L}_{Z_\zeta}e^{a}\,_{\mu}=-\,\lambda^{a}_{\zeta}{}_{b}e^{b}\,_{\mu} \,,\quad
		\mathcal{L}_{Z_\zeta}\omega^{a}\,_{b\mu}=\partial_{\mu}\lambda^{a}_{\zeta}{}_{b}+\omega^{a}\,_{c\mu}\lambda^{c}_{\zeta}{}_{b}-\omega^{c}\,_{b\mu}\lambda^{a}_{\zeta}{}_{c}\,,
	\end{align}
	where $\lambda^{a}_{\zeta}{}_{b}$ defines the Lie algebra homomorphism mapping the symmetry algebra of the vector fields $Z_\zeta$ into the Lorentz algebra. The above equations therefore take the same role in the teleparallel geometry as the Killing equations in Riemannian geometry, while taking into account that the tetrad and spin connection are defined only up to a local Lorentz transformation. The intuitive picture behind this definition is that any given solution to the symmetry condition \eqref{eq:tpsymm}, consists of a tetrad and a spin connection which change along the flow of the symmetry generating vector fields by a local Lorentz transformation only. This ensures that the resulting metric-affine geometry defined by the metric and teleparallel affine connection becomes invariant. Due to the local Lorentz invariance, and thus the freedom to choose a Lorentz transformation at any point, one may always choose a particular gauge, which then enters as an additional condition on the solution alongside with the symmetry condition, so that the symmetry of the remaining gauge-fixed field variables becomes less obvious. For the case of spherical symmetry (i.e., invariant under the group $\mathrm{SO(3)}$), one finds that the most general tetrad in the Weitzenb\"{o}ck gauge $\omega^a{}_{b\mu} \equiv 0$ is given by
	\begin{subequations}\label{sphtetrad}
		\begin{align}
			e^0&=C_1\,\dd t + C_2\, \dd r\,,\\
			e^1&=C_3 \sin\vartheta \cos\varphi \, \dd t + C_4 \sin\vartheta \cos\varphi\,\dd r + (C_5 \cos\vartheta \cos \varphi - C_6 \sin\varphi)\,\dd\vartheta  -\sin\vartheta (C_5 \sin\varphi + C_6 \cos\vartheta \cos\varphi) \,\dd \varphi\,, \\
			e^2&=C_3 \sin\vartheta \sin\varphi\,\dd t + C_4 \sin\vartheta \sin\varphi \,\dd r + (C_5 \cos\vartheta \sin \varphi + C_6 \cos\varphi)\,\dd \vartheta  + \sin\vartheta (C_5 \cos\varphi - C_6 \cos\vartheta \sin\varphi) \,\dd \varphi\,,\\
			e^3&=C_3 \cos\vartheta\,\dd t + C_4 \cos\vartheta\,\dd r - C_5 \sin\vartheta \,\dd \vartheta+ C_6 \sin^2\vartheta\,\dd \varphi,
		\end{align}
	\end{subequations}
	in the usual spherical coordinates $(t, r, \vartheta, \varphi)$, where $C_i=C_i(t,r)$ are six unknown functions depending on the radial and time coordinates. By using Eqs.~\eqref{sphtetrad}, the line element takes the form
	\begin{equation}\label{metric}
		\dd s^2=\left(C_3^2-C_1^2\right)\dd t^2 +2 ( C_3 C_4-C_1 C_2)\,\dd t\,\dd r + \left(C_4^2-C_2^2\right)\dd r^2+ \left(C_5^2+C_6^2\right)\dd \Omega^2\,,
	\end{equation}
	where \(\dd\Omega^2 = \dd\vartheta^2 + \sin^2\vartheta\dd\varphi^2\) is the standard line element on the two-sphere. It is then possible to choose a coordinate system such that the crossed term in the line element cancels out. One simple way of doing that is by introducing the following parametrization for the tetrad functions:
	\begin{subequations}
		\begin{align}
			C_1(t,r)&= A(t,r) \cosh\beta(t,r)\,,& C_3(t,r)&= A(t,r) \sinh\beta(t,r)\,,\\
			C_4(t,r)&= B(t,r) \cosh\beta(t,r)\,,& C_2(t,r)&= B(t,r) \sinh\beta(t,r)\,,\\
			C_5(t,r)&= R(t,r) \cos\alpha(t,r)\,,& C_6(t,r)&= R(t,r) \sin\alpha(t,r)\,,
		\end{align}
	\end{subequations}
	which implies that the line element in Eq.~\eqref{metric} reduces to
	\begin{equation}\label{metric2}
		\dd s^2= - A(t,r)^2\dd t^2  + B(t,r)^2\dd r^2 + R(t,r)^2\dd \Omega^2\,.
	\end{equation}
	Furthermore, the function $R(t,r)$ can be set to $R(t,r)=r$ by an additional coordinate freedom. The final form of the tetrad is thus
	\begin{subequations}\label{tetradF}
		\begin{align}
			e^0&=A\cosh \beta\,\dd t   +   B \sinh \beta \,\dd r\,,\\
			e^1&= A \sin \vartheta \cos \varphi  \sinh \beta \,\dd t +  B \sin \vartheta \cos \varphi  \cosh \beta \,\dd r+ r   (\cos \vartheta \cos \varphi \cos \alpha-\sin \varphi \sin \alpha)\,\dd \vartheta \nonumber\\
			&\phantom{=}-r   \sin \vartheta (\cos \vartheta \cos \varphi \sin \alpha+\sin \varphi \cos \alpha)\,\dd \varphi\,, \\
			e^2&= A\sin \vartheta \sin \varphi  \sinh \beta \,\dd t+  B \sin \vartheta \sin \varphi  \cosh \beta \,\dd r+ r   (\cos \vartheta \sin \varphi \cos \alpha+\cos \varphi \sin \alpha)\,\dd \vartheta \nonumber\\
			&\phantom{=}+ r   \sin \vartheta (\cos \varphi \cos \alpha-\cos \vartheta \sin \varphi \sin \alpha) \,\dd \varphi\,,\\
			e^3&=  A\cos \vartheta  \sinh \beta \,\dd t+   B\cos \vartheta  \cosh \beta \,\dd r  -r   \sin \vartheta \cos \alpha \,\dd \vartheta+ r   \sin ^2\vartheta \sin \alpha\,\dd \varphi \,,
		\end{align}
	\end{subequations}
	which is used in the forthcoming sections to find spherically symmetric solutions. One notices that the metric contains two d.o.f. given by $A(t,r)$ and $B(t,r)$, and the tetrad contains four d.o.f. expressed by $A(t,r)$, $B(t,r)$, $\alpha(t,r)$, and $\beta(t,r)$.

	\subsection{The antisymmetric part of the field equations}\label{sec:asym}

	The field equations of 1-parameter NGR expressed by Eq.~\eqref{fieldeqs} contain symmetric and antisymmetric contributions, where the antisymmetric part of the matter side of the field equations vanishes identically due to the symmetries of the energy-momentum tensor. Hence, for any solution to the field equations, one must impose that the antisymmetric part of the gravitational side also vanishes. The antisymmetric part in spherical symmetry with the tetrad in Eq.~\eqref{tetradF} vanishes identically due to these symmetries except for the two following components
	\begin{align}
		E_{[tr]}&\propto \epsilon   \sin \alpha \left(  B \alpha_{,t} \cosh \beta -  A \alpha_{,r} \sinh \beta \right)=0\,,\label{anti1}\\
		E_{[\vartheta \varphi]}&\propto\epsilon \Big[r A^2 B \Big\{r A_{,r} \alpha_{,r}-2     B \sin \alpha \Big(  \left(B \beta_{,t}-    A_{,r}\right)\cosh \beta+B_{,t} \sinh \beta \Big)\Big\}-r^2 A B^2 \left(B_{,t} \alpha_{,t}+B \alpha_{,tt}\right)\nonumber\\
		&\phantom{=}+r^2 A_{,t} B^3 \alpha_{,t}+  A^3 \Big\{r   \Big(\left(2 B-r B_{,r}\right) \alpha_{,r}+r B \alpha_{,rr}\Big)+2   B^2  \left(r \beta_{,r} \sinh \beta +\cosh \beta \right)\sin \alpha -2   B^3 \sin (2 \alpha )\Big\}\Big]=0\,, \nonumber \\
		\label{anti2}
	\end{align}
	where commas denote derivatives with respect to either $t$ or $r$.
	There are several alternative methods to solve the, each yielding different branches of solutions. In particular, we consider the following three main branches:
	\begin{enumerate}
		\item Branch 1: $\sin\alpha=0$ which implies $\alpha=k \pi$ $(k\in\mathbb{Z})$.
		\item Branch 2: $\sin\alpha\neq 0$ with $\alpha=\alpha_0$ (constant).
		\item Branch 3: $\sin\alpha\neq 0$ with $\alpha=\alpha\left(r,t\right)$ variable and \(B\alpha_{,t}\cosh\beta -  A\alpha_{,r}\sinh\beta = 0\).
	\end{enumerate}
	In~\cite{Golovnev:2023uqb}, the first branch specified above was analyzed in the context of NGR, but the other two branches were omitted in that study. In the Sec. \ref{sec:sph_sol} we analyze all of the branches separately.

	\subsection{The symmetric part of the field equations}\label{sec:sym}
	For the previously identified three branches of solutions, which solve the antisymmetric field equations, it is now possible to derive and display the symmetric field equations. The branches provide three different sets of equations, and so lead to different dynamics and different solutions.

\begin{enumerate}
	\item  Branch 1: The condition $\alpha=k \pi$ $(k\in\mathbb{Z})$ solves both antisymmetric equations in Eqs.~\eqref{anti1}-\eqref{anti2}. Furthermore, it is possible to set $k=0$ without loss of generality. Then, by introducing these conditions into the symmetric part of the field equations in Eq.~\eqref{fieldeqs} we find the following symmetric field equations:
	\begin{align}\label{sym1}
		E_{tt}&= \frac{2 A^2 B_{,r}}{r B^3}-\frac{A^2}{r^2 B^2}+\frac{A^2}{r^2}=\Theta_{tt}\,,\\
		E_{rr}&=\frac{2 A_{,r}}{r A}-\frac{B^2}{r^2}+\frac{1}{r^2}=\Theta_{rr}\,,\label{sym2}\\
		E_{\vartheta\vartheta}&=\frac{r^2 A_{,t} B_{,t}}{A^3 B}-\frac{r^2 A_{,r} B_{,r}}{A B^3}+\frac{r^2 A_{,rr}}{A B^2}+\frac{r A_{,r}}{A B^2}-\frac{r^2 B_{,tt}}{A^2 B}-\frac{r B_{,r}}{B^3}=\Theta_{\vartheta\vartheta}\,,\label{sym3}\\
		E_{tr}&=\frac{2 B_{,t}}{r B}=\Theta_{tr}\,. \label{sym4}
	\end{align}
	Note that in this branch, $\epsilon$ does not appear, meaning that the resulting equations are identical to the GR equations in spherical symmetry. Hence all spherically symmetric solutions to the usual Einstein equations, are solutions of 1PNGR.

	\item  Branch 2: When $\alpha(t,r)=\alpha_0$ with $\alpha_0\neq k \pi$ $(k\in\mathbb{Z})$, the first antisymmetric field equation, Eq.~\eqref{anti1}, is identically satisfied. Thus, one is left with a single antisymmetric field equation (see Eq.~\eqref{anti2}). The symmetric field equations for this branch are as follows:
	\begin{align}
		E_{tt}&= \frac{2 A^2 B_{,r}}{r B^3}-\frac{A^2}{r^2 B^2}-\frac{4 \epsilon \cos (2 \alpha_0) A^2}{9 r^2}+\frac{4 \epsilon A^2}{9 r^2}+\frac{A^2}{r^2}=\Theta_{tt}\,,\label{sym1B}\\
		E_{rr}&=\frac{2 A_{,r}}{r A}+\frac{4 \epsilon \cos (2 \alpha_0) B^2}{9 r^2}-\frac{4 \epsilon B^2}{9 r^2}-\frac{B^2}{r^2}+\frac{1}{r^2}=\Theta_{rr}\,,\label{sym2B}\\
		E_{\vartheta\vartheta}&=\frac{r^2 A_{,t} B_{,t}}{A^3 B}-\frac{r^2 A_{,r} B_{,r}}{A B^3}+\frac{r^2 A_{,rr}}{A B^2}+\frac{r A_{,r}}{A B^2}-\frac{r^2 B_{,tt}}{A^2 B}-\frac{r B_{,r}}{B^3}=\Theta_{\vartheta\vartheta}\,,\label{sym3B}\\
		E_{tr}&=\frac{2 B_{,t}}{r B}=\Theta_{tr}\,.\label{sym4B}
	\end{align}
	These equations feature corrections induced by the teleparallel modifications of general relativity, which are controlled by the constant $\epsilon$. One can also verify that the off-diagonal part of the field equation, $E_{tr}$, coincides with the one from Branch 1.

	\item  Branch 3: This branch is the most complex one since $\alpha=\alpha(t,r)$ with $\sin\alpha\neq0$ implies that the first term in Eq.~\eqref{anti1} is non-zero. Therefore, the only possible way of solving the first antisymmetric field equation is by imposing $ B \alpha_{,t} \cosh \beta -  A \alpha_{,r} \sinh \beta=0$, which introduces complexity in the remaining equations. For this branch, the symmetric field equations become:
	\begin{align}
		E_{tt}&= \frac{2 A^2 B_{,r}}{r B^3}-\frac{A^2}{r^2 B^2}+\frac{8 \epsilon A^2 \alpha_{,r} \sin (\alpha) \cosh (\beta)}{9 r B}+\frac{2 \epsilon A^2 \alpha_{,r}^2}{9 B^2}-\frac{4 \epsilon A^2 \cos (2 \alpha)}{9 r^2}+\frac{4 \epsilon A^2}{9 r^2}+\frac{A^2}{r^2}+\frac{2}{9} \epsilon \alpha_{,t}^2=\Theta_{tt}\,,\label{sym1C}\\
		E_{rr}&=\frac{2 A_{,r}}{r A}+\frac{8 \epsilon B^2 \alpha_{,t} \sin (\alpha) \sinh (\beta)}{9 r A}+\frac{2 \epsilon B^2 \alpha_{,t}^2}{9 A^2}+\frac{4 \epsilon B^2 \cos (2 \alpha)}{9 r^2}-\frac{4 \epsilon B^2}{9 r^2}-\frac{B^2}{r^2}+\frac{2}{9} \epsilon \alpha_{,r}^2+\frac{1}{r^2}=\Theta_{rr}\,,\label{sym2C}\\
		E_{\vartheta\vartheta}&=\frac{r^2 A_{,t} B_{,t}}{A^3 B}-\frac{r^2 A_{,r} B_{,r}}{A B^3}+\frac{r^2 A_{,rr}}{A B^2}+\frac{r A_{,r}}{A B^2}-\frac{r^2 B_{,tt}}{A^2 B}+\frac{2 \epsilon r^2 \alpha_{,t}^2}{9 A^2}+\frac{4 \epsilon r \alpha_{,t} \sin (\alpha) \sinh (\beta)}{9 A}-\frac{r B_{,r}}{B^3}\nonumber\\
		&\phantom{=}-\frac{2 \epsilon r^2 \alpha_{,r}^2}{9 B^2}-\frac{4 \epsilon r \alpha_{,r} \sin (\alpha) \cosh (\beta)}{9 B}=\Theta_{\vartheta\vartheta}\,,\label{sym3C}\\
		E_{tr}&=\frac{2 B_{,t}}{r B}+\frac{8 \epsilon B \alpha_{,t} \sin (\alpha) \cosh (\beta)}{9 r}+\frac{4}{9} \epsilon \alpha_{,r} \alpha_{,t}=\Theta_{tr}\,,\label{sym4C}
	\end{align}
	which form indeed a more complicated set of differential equations in comparison with the ones obtained in Branches 1 and 2.
\end{enumerate}
Having found all field equations, we solve the vacuum solutions of the equations for each branch next.

	\section{Spherical symmetric vacuum solutions}\label{sec:sph_sol}
	In order to find the spherically symmetric vacuum solutions of the theory, we set the energy momentum tensor to zero, and study the symmetric field equations for each branch that we identified from the antisymmetric field equations in section \ref{sec:asym}.

	\subsection{Branch 1: $\sin\alpha=0$ }\label{sssec:br1}
	 We now solve the system of differential equations expressed in Eqs.~\eqref{sym1}-\eqref{sym4} for the vacuum case. From Eq.~\eqref{sym4} we find that the $tr$ component takes the following constraint form
	\begin{equation}
		E_{(tr)}\propto B_{,t}=0\,,\quad B(t,r)=B(r)\,.
	\end{equation}
	Furthermore, by using the $tt$ and $rr$ components of the field equations (see Eqs.~\eqref{sym1} and~\eqref{sym2}) one immediately finds
	\begin{equation}
		A(t,r)^2=\Big(1-\frac{2M}{r}\Big)A_0(t)\,,\quad B(r)^2=\Big(1-\frac{2M}{r}\Big)^{-1}\,,
	\end{equation}
	where $A_0(t)$ is an integration function and $M$ is a constant that plays the role of the mass. The above form of the metric functions corresponds to the Schwarzschild metric. This can be clarified by performing a time-coordinate transformation $A_0(t)\dd t\rightarrow \dd t$ such that one absorbs the arbitrary function $A_0(t)$. It is worth mentioning that the tetrad in Eq.~\eqref{tetradF} for this branch
	\begin{align}
		e^0&=\Big(1-\frac{2M}{r}\Big)^{1/2}\cosh \beta\,\dd t +   \Big(1-\frac{2M}{r}\Big)^{-1/2}\sinh \beta \,\dd r\,,  \\
		e^1&=   \Big(1-\frac{2M}{r}\Big)^{1/2}\sin \vartheta \cos \varphi  \sinh \beta\,\dd t +  \Big(1-\frac{2M}{r}\Big)^{-1/2} \sin \vartheta \cos \varphi  \cosh \beta \,\dd r+ r   \cos \vartheta \cos \varphi\,\dd \vartheta \nonumber\\
		&\phantom{=}-r   \sin \vartheta \sin \varphi\,\dd \varphi \\
		e^2&= \Big(1-\frac{2M}{r}\Big)^{1/2} \sin \vartheta \sin \varphi \sinh \beta \,\dd t+  \Big(1-\frac{2M}{r}\Big)^{-1/2} \sin \vartheta \sin \varphi  \cosh \beta \,\dd r + r   \cos \vartheta \sin \varphi \,\dd \vartheta \nonumber\\
		&\phantom{=}+ r   \sin \vartheta \cos \varphi \,\dd \varphi \\
		e^3&= \Big(1-\frac{2M}{r}\Big)^{1/2} \cos \vartheta  \sinh \beta\,\dd t +  \Big(1-\frac{2M}{r}\Big)^{-1/2}  \cos \vartheta \cosh \beta \,\dd r  -r   \sin \vartheta \,\dd \vartheta   \,,
	\end{align}
	contains $\beta=\beta(t,r)$ as any arbitrary function. This computation showed that, on the level of the metric there is a unique spherically symmetric solution for the Branch 1 in 1-parameter NGR  which is described by the Schwarzschild metric. This is obvious to obtain since the symmetric field equations in Eqs.~\eqref{sym1}-\eqref{sym4} are identical to the GR ones, while the antisymmetric field equations are solved, leading $\beta$ as an arbitrary tetrad function. This means that the tetrad is not unique, since it contains the free function $\beta$. In principle, this could have an impact at the perturbation level.

	\subsection{Branch 2: $\sin\alpha\neq 0$  with $\alpha=\alpha_0$ (constant)}\label{sssec:br2}
	Similarly as we computed it in the previous section, one can use the $tt,$ $rr$ and $tr,$ components of the field equations (see Eqs. \eqref{sym1B}, \eqref{sym2B} and \eqref{sym4B} respectively) to obtain
	\begin{equation}
		A(t,r)^2=\Big(1-\frac{2M}{r}\Big)A_0(t)\,,\quad B(r)^2=h\Big(1-\frac{2M}{r}\Big)^{-1}\,,
	\end{equation}
	where again one notices that one can re-define the time-coordinate such that the time function $A_0(t)$ is absorbed. Here, for the sake of notation simplicity, we have introduced the parameter
	\begin{equation}
		h=h(\alpha,\epsilon)=\frac{1}{1-\frac{4}{9}\epsilon\left[\cos\left(2\alpha\right)-1\right]}\,.
	\end{equation}
	Then, for this branch, the metric must be  static in the absence of matter fields, and the solution is described by a modified Schwarzschild-like form with the following line-element
	\begin{equation}\label{eq:metric0}
		\dd s^2=-\Big(1-\frac{2M}{r}\Big)\dd t^2+h\Big(1-\frac{2M}{r}\Big)^{-1}\dd r^2+r^2\dd\Omega^2\,.
	\end{equation}
	There is still one remaining field equation in the system, Eq.~\eqref{anti2}, which under the line element above takes the form
	\begin{equation}
		0=\epsilon \Big[\frac{3   r (2 M-r) \beta_{,r} \sinh \beta }{B}+\frac{3   (M-r) \cosh \beta }{B}+3   r^2 \sqrt{1-\frac{2 M}{r}} \beta_{,t} \cosh \beta +6   \cos\alpha_0 (r-2 M)\Big]\,,
	\end{equation}
	which constrains the form of the tetrad function $\beta(t,r)$ and does not affect the form of the metric. In general, the above differential equation cannot be easily solved. However, for the specific case when $\beta(t,r)=\beta(r)$ we obtain the analytical solution
	\begin{equation}\label{eq:betaII}
		\beta(r)=\arccosh\Big[\frac{2    B(r)\cos \alpha_0    }{r}\left(M^2 \beta_0+r\right)\Big]\,,
	\end{equation}
	for which the tetrad in the static case reduces to
	\begin{align}
	e^0&=\frac{2 \sqrt{h} \cos \alpha_0 \left(\beta_0 M^2+r\right)}{r}\,\dd t+\sqrt{\frac{h}{r-2 M}} \sqrt{\frac{4 h \cos ^2\alpha_0 \left(\beta_0 M^2+r\right)^2}{r-2 M}-r}\,\dd r\,,   \\
	e^1&= \frac{\sin \vartheta \cos \varphi \sqrt{4 h \cos ^2\alpha_0 \left(\beta_0 M^2+r\right)^2+r (2 M-r)}}{r}\dd t-\frac{2 h \cos \alpha_0 \sin \vartheta \cos \varphi \left(\beta_0 M^2+r\right)}{2 M-r}\,\dd r \nonumber\\
	&\phantom{=}+r (\cos \alpha_0 \cos \vartheta  \cos \varphi -\sin \alpha_0 \sin \varphi )\dd\vartheta-r \sin \vartheta  (\sin \alpha_0 \cos \vartheta  \cos \varphi +\cos \alpha_0 \sin \varphi )\dd \varphi\,, \\
	e^2&=\frac{\sin \vartheta \sin \varphi \sqrt{4 h \cos ^2\alpha_0 \left(\beta_0 M^2+r\right)^2+r (2 M-r)}}{r}\,\dd t-\frac{2 h \cos \alpha_0 \sin \vartheta \sin \varphi \left(\beta_0 M^2+r\right)}{2 M-r}\,\dd r \nonumber\\
	&\phantom{=}+r (\cos \alpha_0 \cos \vartheta  \sin \varphi +\sin \alpha_0\cos \varphi )\dd\vartheta+r \sin \vartheta  (\cos \alpha_0 \cos \varphi -\sin \alpha_0 \cos \vartheta  \sin \varphi )\dd\varphi\,, \\
	e^3&=\frac{\cos \vartheta \sqrt{4 h \cos ^2\alpha_0 \left(\beta_0 M^2+r\right)^2+r (2 M-r)}}{r}\dd t-\frac{2 h \cos \alpha_0 \cos \vartheta \left(\beta_0 M^2+r\right)}{2 M-r}\,\dd r\nonumber\\
	&\phantom{=}-r \cos \alpha_0 \sin \vartheta\, \dd\vartheta+r \sin \alpha_0\sin ^2\vartheta\,\dd\varphi\,.
\end{align}
It is worth mentioning that, for the specific case when $\alpha_0=\pi/2$, the form of the tetrad reduces to the complex tetrad form used in~\cite{Bahamonde:2021srr,Bahamonde:2022lvh,Bahamonde:2022chq}, where exact and numerical black holes solutions were found in $f(T)$ gravity and also in scalar-tensor theories. Note, however, that this tetrad is not necessarily complex for \(r > 2M\), provided that the parameters of the model satisfy the conditions \(h > 0\), \(4h\cos^2\alpha_0 > 1\), and the following quadratic equation
\begin{equation}
4 h \cos ^2\alpha_0 \left(\beta_0 M^2+r\right)^2+r (2 M-r) = 0
\end{equation}
does not have any solutions in the region \(r > 2M\). On the other hand, in the region \(r < 2M\), where \(e^0\) becomes spacelike and \(e^1\) becomes timelike, their roles must be reversed in the ansatz given in Eq.~\eqref{tetradF}, such that the interior solution is kept real as well. We like to point out that this reversal is only due to the choice of coordinates and the Lorentz gauge, while the metric and torsion defining the teleparallel connection remain real.

	\subsection{Branch 3: $\sin\alpha\neq 0$  with $\alpha=\alpha\left(r,t\right)$ variable}\label{sssec:br3}
	This branch is more involved than the previous ones (see Eqs.~\eqref{sym1C}-\eqref{sym4C} for the vacuum case). In order to explore this branch, let us assume for the entirety of this section that the functions $B(r,t)$ and $A(r,t)$ satisfy the constraint $B=1/A$ (i.e. $g_{rr}=-1/g_{tt}$), a choice for which the equations simplify significantly. Under this assumption, the system of Eqs.~\eqref{sym1C}-\eqref{sym4C} with the antisymmetric Eqs.~\eqref{anti1}-\eqref{anti2} can be rewritten by using Eq.~\eqref{anti1}, yielding
	\begin{align}
		\alpha_{,r}&=-\frac{2     \sin \alpha  \cosh \beta }{r A}\,,\quad  \alpha_{,t}=-\frac{2     A\sin \alpha  \sinh \beta }{r}\,,\label{eqA0}\\
		0&=\sinh \beta \Big[  A\Big(\left(A-r A_{,r}\right) \tanh \beta -r A \beta_{,r} \Big)+  r \beta_{,t}  \tanh \beta \Big]\,,\label{eqA1}\\
		0&=2 r A_{,r} A+A^2-1\,,\quad A_{,t}=0\,.\label{eqA}
	\end{align}
	Equation ~\eqref{eqA} can be directly integrated in order to obtain $A(t,r)^2=1-2M/r$. This means that there is a unique solution for the metric functions when we assume $B=1/A$, which is given by the Schwarzschild spacetime. This conclusion holds for the generic case where the functions depend on $r$ and $t$. Consequently, any solution beyond Schwarzschild must satisfy $B\neq1/A$. Even though we already  found the form of the metric, Eqs~\eqref{eqA0}-\eqref{eqA1} need to be solved for $\alpha(t,r)$ and $\beta(t,r)$. Solving the equations above proves to be a difficult task. Nevertheless, taking the assumption $\beta(t,r)=\beta(r)$ one finds
	\begin{equation}
		\beta(r)=\arcsinh\Big[\frac{c_1 r^{3/2}}{\sqrt{r-2M}}\Big]\,,\quad \alpha(t,r)=2 \arccot\Big[e^{2 c_1     t +\alpha_2(r)}\Big]\,,\quad \alpha_2(r)=\int dr \frac{2   \sqrt{c_1^2 r^2+1-2M/r}}{r-2M}\,,
	\end{equation}
	yielding the following form of the tetrad
	\begin{align}
		e^0&= \sqrt{1-\frac{2 M}{r}+c_1^2 r^2} \,\dd t+ \frac{c_1 r^2}{r-2 M}\,\dd r\,,  \\
		e^1&= c_1 r \sin \vartheta \cos \varphi \,\dd t+ \frac{\sqrt{r \left(c_1^2 r^3+r-2 M\right)}}{r-2 M}\sin \vartheta \cos \varphi \,\dd r + r \Big(\cos \vartheta \cos \varphi \sinh (2 c_1 t+\alpha_2)-\sin \varphi\Big)\sech(2 c_1 t+\alpha_2)\,\dd \vartheta \nonumber\\
		&\phantom{=}-r \sin \vartheta\Big(\sin \varphi \sinh (2 c_1 t+\alpha_2)+\cos \vartheta \cos \varphi\Big) \sech(2 c_1 t+\alpha_2) \,\dd \varphi\,,\\
		e^2&= c_1 r \sin \vartheta \sin \varphi \,\dd t+ \frac{\sin \vartheta \sin \varphi \sqrt{r \left(c_1^2 r^3+r-2 M\right)}}{r-2 M}\,\dd r + r \Big(\cos \vartheta \sin \varphi \sinh (2 c_1 t+\alpha_2)+\cos \varphi\Big)\sech(2 c_1 t+\alpha_2) \,\dd \vartheta\nonumber\\
		&\phantom{=}+ r \sin \vartheta \sech(2 c_1 t+\alpha_2) (\cos \varphi \sinh (2 c_1 t+\alpha_2)-\cos \vartheta \sin \varphi)\,\dd \varphi\,, \\
		e^3&= c_1 r \cos \vartheta\,\dd t + \frac{\cos \vartheta \sqrt{r \left(c_1^2 r^3-2 M+r\right)}}{r-2 M}\,\dd r  -r \sin \vartheta \tanh (2 c_1 t+\alpha_2) \,\dd \vartheta+ r \sin ^2\vartheta \sech(2 c_1 t+\alpha_2) \,\dd \varphi\,.
	\end{align}
	This indicates that, even though the metric is static (Schwarzschild), the extra degrees of freedom that only appear in the tetrad are non-static in general.  Thus, one obtains the Schwarzschild metric, with the Minkowski limit \(M \to 0\), albeit expressed by a boosted tetrad with a non-trivial coordinate dependent \(\alpha(t,r)\).

	For the static case, one sets $c_1=0$, and this implies consequently that $\beta=0$ and that the integral appearing in $\alpha(r)$ can be explicitly obtained yielding
	\begin{equation}
		\alpha_2(r)=\alpha_0+4 \tanh ^{-1}\left(\sqrt{1-\frac{2 M}{r}}\right)\quad \Longrightarrow   \alpha(r)=2 \arccot\left[e^{\alpha_0+4 \arctanh\left(\sqrt{1-\frac{2 M}{r}}\right)}\right]\,,
	\end{equation}
	where $\alpha_0$ is an integration constant. Furthermore, for the non-static case in Minkowski ($M=0$), we find that the integral appearing in $\alpha_2$ in the above equation can be solved yielding the following form for $\alpha_2$:
	\begin{equation}\label{alpha2B}
		\alpha_2(r)= 2 \Big[\sqrt{1+c_1^2 r^2}-\arctanh\left(\sqrt{1+c_1^2 r^2}\right)\Big]+\alpha_0\,,
	\end{equation}
	which is a Minkowski solution for our theory with a non-trivial time and radial dependence tetrad given by
	\begin{align}
		e^0&=\sqrt{1+c_1^2 r^2}\,\dd t + c_1 r \,\dd r\,,  \\
		e^1&= c_1 r \sin \vartheta \cos \varphi\,\dd t + \sqrt{1+c_1^2 r^2} \sin \vartheta \cos \varphi \,\dd r+ \Big[r \cos \vartheta \cos \varphi \tanh (2 c_1 t+\alpha_2)-r \sin \varphi \sech(2 c_1 t+\alpha_2)\Big]\,\dd \vartheta \nonumber\\
		&\phantom{=} -r \sin \vartheta  \Big[\sin \varphi \tanh(2 c_1 t+\alpha_2)+\cos \vartheta \cos \varphi\sech(2 c_1 t+\alpha_2)\Big]\,\dd \varphi\,, \\
		e^2&= c_1 r \sin \vartheta \sin \varphi\,\dd t + \sqrt{1+c_1^2 r^2} \sin \vartheta \sin \varphi \,\dd r+ r  \Big[\cos \vartheta \sin \varphi \tanh (2 c_1 t+\alpha_2)+\sech(2 c_1 t+\alpha_2)\cos \varphi\Big]\,\dd \vartheta\nonumber\\
		&\phantom{=}+ r \sin \vartheta \Big[\cos \varphi \tanh (2 c_1 t+\alpha_2)-\cos \vartheta \sin \varphi \sech(2 c_1 t+\alpha_2)\Big] \,\dd \varphi\,,\\
		e^3&= c_1 r \cos \vartheta\,\dd t + \sqrt{1+c_1^2 r^2} \cos \vartheta \,\dd r -r \sin \vartheta \tanh (2 c_1 t+\alpha_2) \,\dd \vartheta+ r \sin ^2\vartheta \sech(2 c_1 t+\alpha_2)\,\dd \varphi\,,
	\end{align}
	with $\alpha_2$ given by Eq.~\eqref{alpha2B}.

	\section{Observables in spherically symmetric vacuum solutions}\label{sec:Obs}

	Our analysis of the 1PNGR field equations implies that there exist spherically symmetric solutions beyond Schwarzschild geometry. For the derivation of the observable consequences of such findings, and to be able to identify constraints on the 1PNGR parameter $\epsilon$, we focus on the analytic solution we found in Branch 2, see Sec. \ref{sssec:br2}. These solutions contain a second parameter $\alpha=\alpha_0$, which emerges from solving the field equations. In this section, we study the properties of this new vacuum solution. In section~\ref{ssec:geomprop}, we start by studying its geometric properties, such as horizons and asymptotic flatness. In section~\ref{ssec:basobs}, we study the classical observables, in particular the perihelion shift, light deflection and Shapiro delay. In section~\ref{ssec:bhobs} we discuss the Komar mass and singularities. Finally, in section~\ref{ssec:shadow} we study the photon sphere and shadow of the solution.

	\subsection{Geometric properties of the solution}\label{ssec:geomprop}

The metric for which we perform the following phenomenological analysis is described by the line-element given in Eq. \eqref{eq:metric0}, which can be recast into the more convenient form
	\begin{equation}\label{eq:metric}
		\dd s^2=-f\left(r\right)\dd t^2+\frac{h\left(\alpha,\epsilon\right)}{f\left(r\right)}\dd r^2+r^2\dd\Omega^2,
	\end{equation}
	where the functions $f\left(r\right)$ and $h\left(\alpha,\epsilon\right)$ are given explicitly by
	\begin{equation}
		f\left(r\right)=1-\frac{2M}{r}\,,\quad  h\left(\alpha,\epsilon\right) = \Big(1-\frac{4}{9}\epsilon\left[\cos\left(2\alpha\right)-1\right]\Big)^{-1}\,.
	\end{equation}
Before proceeding with the analysis that follows, a few considerations are worth mentioning:
	\begin{itemize}
		\item To preserve the Lorentzian signature of the metric, the condition $ h\left(\alpha,\epsilon\right)>0$ must hold for all values of $\epsilon$ and $\alpha$. Then, the metric
		features an event horizon at $r_H=2M$, at which both the $g_{tt}$ and $g_{rr}$ components of the metric change signs. For a solution with an arbitrary $\alpha$ and $\epsilon$ to preserve the same signs of the metric components as the Schwarzschild solution ($\epsilon=0$), it is necessary that the constant $\epsilon$ satisfies the following requirement
		\begin{equation}
			\epsilon>\frac{9}{4}\left[\cos\left(2\alpha\right)-1\right]^{-1}\equiv\delta_{min}<0.
		\end{equation}
		\item The solution is asymptotically non-flat. The Ricci tensor of the metric features the following non-vanishing components
			\begin{equation}
				\mathring{R}_{\varphi\varphi} = \mathring{R}_{\vartheta\vartheta} \sin^2\vartheta =\left(1 -  \frac{1}{h\left(\alpha,\epsilon\right)}\right) \sin^2\vartheta\,.
		\end{equation}
		\item Due to the spherical symmetry of the solution, the motion of particles in the spacetime under consideration is fully determined by the normalization condition for curves $g_{\mu\nu}( x)\dot x^\mu \dot x^\nu = \sigma$ restricted to the equatorial plane $\vartheta=\pi/2$, where $\sigma=0$ for null and $\sigma = -1$ for timelike particle trajectories. This equation can be rewritten in the form
		\begin{equation}
			\frac{1}{2}\dot r^2 + V_{\textrm{eff}}(r,E,L,\sigma) = 0\,,
		\end{equation}
		where here, the effective potential, in terms of the constants of motion $E$ and $L$ (energy and angular momentum, respectively), is given by
		\begin{equation}
			V_{\textrm{eff}}(r) = \frac{1}{h(\alpha,\epsilon)} \left[- \frac{E^2}{2} + \frac{f}{2 } \left(\frac{L^2}{r^2} -  \sigma\right) \right] \,,\quad  E = -  f \dot t\,,\quad L =  r^2 \dot{\varphi}\,.
		\end{equation}
		Thus, the teleparallel modification of the effective potential in comparison with the Schwarzschild result is given by $V_{\textrm{eff}} = h(\epsilon, \alpha)^{-1} V_{\textrm{effSchw}}$, where the subscript $_{\rm Schw}$ denotes the Schwarzschild counterpart of the same quantity.
	\end{itemize}

	In general, for particle motion, this implies that the orbits of test particles in the 1PNGR generalization of the Schwarzschild spacetime differ from the ones Schwarzschild geometry only by a constant factor depending on the teleparallel parameters $\epsilon$ and $\alpha$, since
	\begin{align}
		\dot r = \frac{1}{\sqrt{h(\alpha,\epsilon)}} \dot r_{\textrm{Schw}}\,,\quad
		\frac{\dd\varphi}{\dd r} = \frac{\dot \varphi}{\dot r}  =\sqrt{h(\alpha,\epsilon)} \frac{\dd\varphi}{\dd r}_{\textrm{Schw}}\,,\quad
		\frac{\dd t}{\dd r}      = \frac{\dot t}{\dot r}  =\sqrt{h(\alpha,\epsilon)} \frac{\dd t}{\dd r}_{\textrm{Schw}}\,.
	\end{align}
	In particular, the circular orbits, which are characterized by the conditions $\dot r=\ddot r = 0$, are identical as in Schwarzschild spacetime.

	\subsection{Classical observables}\label{ssec:basobs}
	For the classical observables, i.e., light deflection $\Delta \varphi_{\textrm{light}}$, perihelion shift $\Delta \varphi_{\textrm{peri}}$, or the Shapiro delay $\Delta t_{\textrm{Shap}}$, which are derived from the different parametrizations of the orbits above, this allows for the following comparison:
	\begin{itemize}
		\item The perihelion shift $\Delta \varphi_{\textrm{peri}}$ is determined from the orbits parametrized as curves $r(\varphi)$ or $\varphi(r)$. If the orbits were perfect elipses, the angular difference between the perihelion $R_-$ and the aphelion $R_+$ would be exactly $\pi$. Any deviations from this value define the perihelion shift $\Delta \varphi$,
		\begin{align}
			\Delta \varphi_{\textrm{peri}}
			&= 2 \left| \int_{R_-}^{R_+} \frac{\dd\varphi}{\dd r}\dd r \right|  - 2\pi\nonumber\\
			&=\sqrt{h(\alpha,\epsilon)}   2\left| \int_{R_-}^{R_+} \frac{\dd\varphi}{\dd r}_{\textrm{Schw}}\dd r \right| - 2\pi\nonumber\\
			&=\sqrt{h(\alpha,\epsilon)}\  \Delta \varphi_{\textrm{periSchw}} + ( \sqrt{h(\alpha,\epsilon)} - 1 )2\pi\,.
		\end{align}
		\item The light deflection by isolated gravitating objects is determined as angle $\Delta \varphi_{\textrm{light}}$, about which a light ray bends when it passes by a gravitating object. Since we are studying the effect on an asymptotically non-flat spacetime, we consider a light ray that is emitted by a source at $r=R_S$ with angle $\Psi_S(R_S) = L/E \sqrt{f(R_S)}/R_S$ with respect to the radial direction, and observed by a receiver at $r=R_R$ with an angle $\Psi_R(R_R)= L/E \sqrt{f(R_R)}/R_R$ with respect to the radial direction. Following \cite{Ishihara:2016vdc}, the deflection angle $\Delta \varphi_{\textrm{light}}$ can be calculated as
		\begin{align}
			\Delta \varphi_{\textrm{light}}
			&= \int_{r_c}^{R_S} \frac{\dd\varphi}{\dd r} \dd r + \int_{r_c}^{R_R} \frac{\dd\varphi}{\dd r} \dd r + \Psi_R - \Psi_S\nonumber\\
			&= \sqrt{h(\alpha,\epsilon)}  \left(\int_{r_c}^{R_S} \frac{\dd\varphi}{\dd r}_{\textrm{Schw}} \dd r + \int_{r_c}^{R_R} \frac{\dd\varphi}{\dd r}_{\textrm{Schw}} \dd r \right)+ \Psi_R{}_{\textrm{Schw}} - \Psi_S{}_{\textrm{Schw}}\nonumber\\
			&= \sqrt{h(\alpha,\epsilon)}\ \Delta \varphi_{\textrm{lightSchw}} + (1 - \sqrt{h(\alpha,\epsilon)} ) (\Psi_R{}_{\textrm{Schw}} - \Psi_S{}_{\textrm{Schw}})\,.
		\end{align}
		The angles $\Psi_S$ and $\Psi_R$ depend solely on the metric component $g_{tt}(r)$, which explains why they take identical values as in Schwarzschild geometry.
		\item The Shapiro time delay is the additional time in the travel time of a light signal in the presence of a gravitating object, compared to the absence of the object. Again, the derivation is straightforward
		\begin{align}
			\Delta t_{\textrm{Shap}}
			&= \int_{r_1}^{r_2} \frac{\dd t}{\dd r} \dd r - \left(\int_{r_1}^{r_2} \frac{\dd t}{\dd r}_{\textrm{Schw}} \dd r\right)_{M=0}\nonumber\\
			&= \sqrt{h(\alpha,\epsilon)} \int_{r_1}^{r_2} \frac{\dd t}{\dd r}_{\textrm{Schw}} \dd r - \left(\int_{r_1}^{r_2} \frac{\dd t}{\dd r}_{\textrm{Schw}} \dd r\right)_{M=0}\nonumber\\
			&= \sqrt{h(\alpha,\epsilon)}\ \Delta t_{\textrm{ShapSchw}}   + (\sqrt{h(\alpha,\epsilon)} - 1)  \left(\int_{r_1}^{r_2} \frac{\dd t}{\dd r}_{\textrm{Schw}} \dd r\right)_{M=0}\,.
		\end{align}
	\end{itemize}

	\subsection{Komar mass and singularities}\label{ssec:bhobs}
		Defining the quantity $\bar\delta = \epsilon /\delta_{min}$, which by construction is constrained to vary in the interval $\bar\delta\in (-\infty, 1]$, with $\bar\delta=0$ corresponding to the Schwarzschild solution, the function $h\left(\alpha,\epsilon\right)$ can be conveniently redefined into a one-parameter function of the form
	\begin{equation}
		h\left(\alpha,\epsilon\right)=h\left(\bar\delta\right)=\frac{1}{1-\bar\delta}.
	\end{equation}
	Indeed, there are an infinite amount of parameter combinations $\alpha$ and $\epsilon$ that result in the same value of $\bar\delta$ and, consequently, correspond to the same spacetime metric. Furthermore, all possible solutions satisfying the requirement $\epsilon>\delta_{min}$ can be mapped to a value of $\bar\delta<1$. Thus, instead of analyzing the parameters $\alpha$ and $\epsilon$ separately in what follows, we chose to analyze different values of $\bar\delta$.

	It is also useful to calculate the Komar mass $\mathcal{M}$ for this spacetime which is related to the force needed by an observer at infinity to keep a spherical uniform mass distribution. For that, one must assume the existence of a timelike Killing vector field $\xi^\mu=\{1,0,0,0\}$ and a spacelike hypersurface $\Sigma_t$ from the event horizon to spatial infinity in a constant slice $t$ whose normal vector is $n_\mu=\{-g_{tt},0,0,0\}$. Then, the Komar mass reads~\cite{Carroll:2004st}
	\begin{equation}
	\mathcal{M}=-\frac{1}{8\pi}\int_{S_t}\lc{\nabla}^\mu \xi^\nu dS_{\mu\nu}\,,
	\end{equation}
	where $S_t$ is the 2-boundary of $\Sigma_t$ and $\dd S_{\mu\nu}$ is the surface element of $S_t$ which is $\dd S_{\mu\nu}=-2n_{[\mu}\sigma_{\nu]}\sqrt{s}\dd\vartheta\dd\varphi$ with $s=r^4\sin^2\vartheta$ being the determinant of the 2-dimensional metric on $S_t$ and $\sigma_\mu=\{0,\sqrt{g_{rr}},0,0\}$. By replacing the form of the metric in Eq.~\eqref{eq:metric} into the equations above, we find that the Komar mass of the spacetime is
	\begin{equation}
		\mathcal{M}=\frac{1}{8\pi}\lim_{r\rightarrow \infty}\int_{0}^{\pi}\int_{0}^{2\pi}\frac{r^2g'_{tt}}{\sqrt{g_{tt}g_{rr}}}\,\sin\vartheta\, \dd\varphi \, \dd \vartheta=M \sqrt{1+\frac{8}{9} \epsilon \sin ^2(\alpha_0)}\,.
	\end{equation}
	It is also useful to compute the Kretschmann invariant $\mathring{K}=\mathring{R}^{\alpha\beta\mu\nu}\mathring{R}_{\alpha\beta\mu\nu}$ for our spacetime whose value becomes
	\begin{equation}
		\mathring{K}=\frac{16 M^2 (-4 \epsilon \cos (2 \alpha_0)+4 \epsilon+9)^2}{27 r^6}+\frac{128 \epsilon M \sin ^2(\alpha_0) (4 \epsilon \cos (2 \alpha_0)-4 \epsilon-9)}{81 r^5}+\frac{256 \epsilon^2 \sin ^4(\alpha_0)}{81 r^4}\,,
	\end{equation}
	from which one observes that, similarly to the Schwarzschild case, the solutions considered features a singularity at $r=0$. Note that for $\epsilon \neq 0$ it is also easy to see this by computing the Ricci scalar which is $\mathring{R}=\frac{2(h-1)}{h r^2}$ where one can also see the singularity appearing at the origin.

	\subsection{Photon sphere and shadow}\label{ssec:shadow}
\begin{figure}
		\includegraphics[scale=0.38]{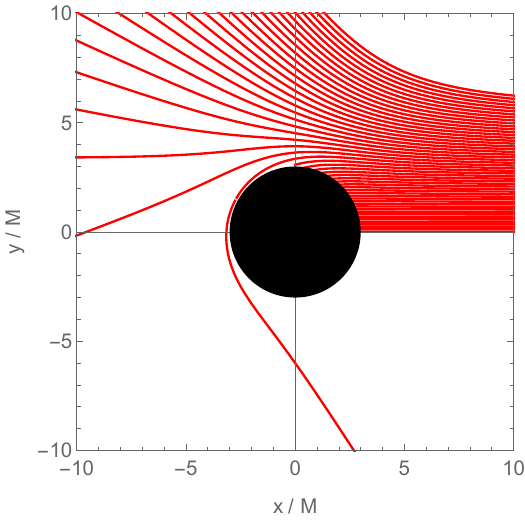}
		\includegraphics[scale=0.38]{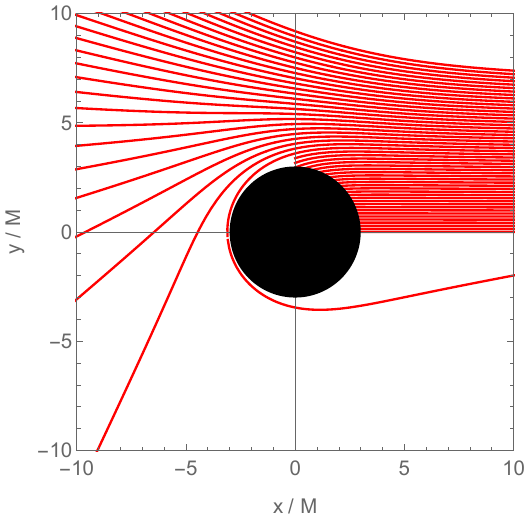}
		\includegraphics[scale=0.38]{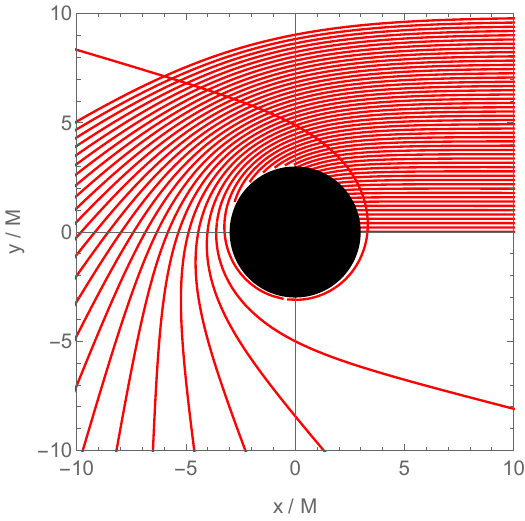}
		\includegraphics[scale=0.38]{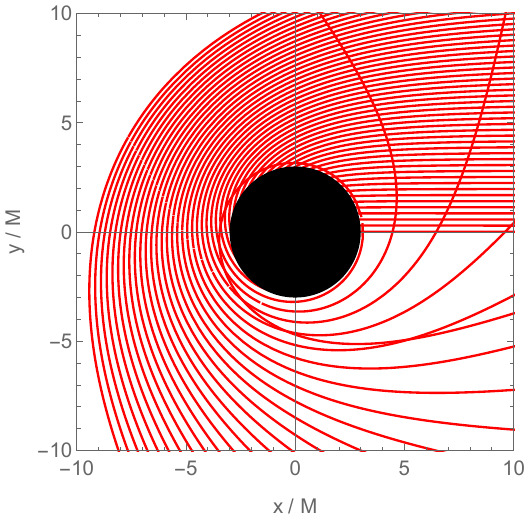}
		\includegraphics[scale=0.38]{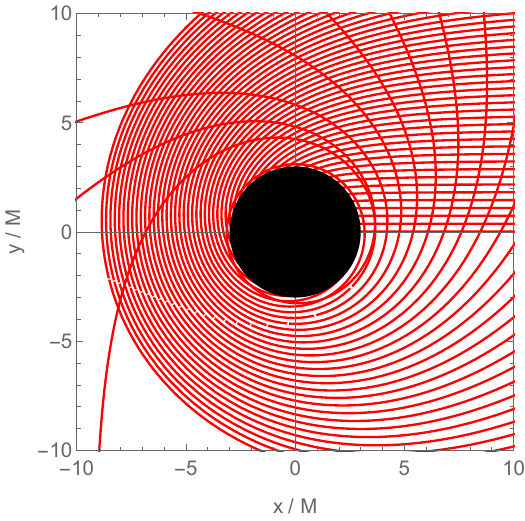}
		\caption{Geodesic congruences for $\bar\delta=\{-2; -1; 0; 0.5; 0.7\}$, from left to right. The black disk represents the photon sphere at $r=3M$, i.e., all photons approaching this disk eventually reach the event horizon.}
		\label{fig:geodesic}
	\end{figure}

	Let us now analyze the observational properties of the spherically symmetric solutions deduced in this manuscript when surrounded by optically-thin accretion disks. To produce the observed images and intensity profiles of the solutions in consideration, we recur to a Mathematica-based ray-tracing code previously used in other publications \cite{Rosa:2023qcv,Rosa:2023hfm,Olmo:2023lil,Rosa:2022tfv,daSilva:2023jxa,Guerrero:2022msp,Guerrero:2022qkh,Olmo:2021piq,Guerrero:2021ues}. In this code, the trajectories of photons are computed via numerical solutions of the geodesic equation. For spherically symmetric solutions like the ones considered in this work, the geodesic equation can be conveniently rewritten in terms of radial derivatives of the azimuthal angle in the form
	\begin{equation}
		\varphi'\left(r\right)=\sqrt{h}\varphi'_{\rm Schw}=\pm \frac{b}{r^2}\frac{\sqrt{-g_{tt}g_{rr}}}{\sqrt{1+g_{tt}\frac{b^2}{r^2}}},
	\end{equation}
	where $b\equiv L/E$ is the impact parameter, with $L$ the angular momentum and $E$ the energy of the photon. In Fig.\ref{fig:geodesic} we provide a set of null geodesics computed in the background spacetime for models with different values of the parameter $\bar\delta$ and different impact parameters $b$. These panels clarify the effect of $\bar\delta$ in the propagation of photons. Indeed, a comparison with the middle panel, corresponding to the Schwarzschild solution, with $\bar\delta=0$, indicates that positive values of $\bar\delta$ have a repulsive effect in the photons approaching the central black-hole, whereas negative values of $\bar\delta$ contribute with an additional attractive effect. These additional effects are expected to induce non-negligible qualitative modifications in the observational properties of the solutions considered in comparison to the Schwarzschild solution.

\begin{figure}
		\includegraphics[scale=0.66]{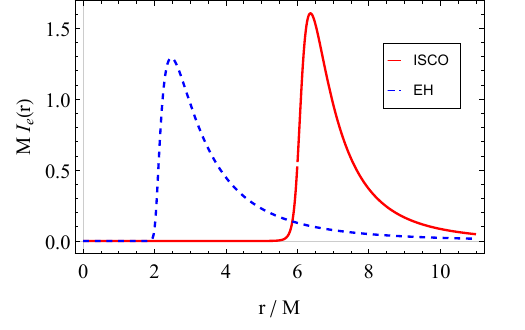}
		\includegraphics[scale=0.65]{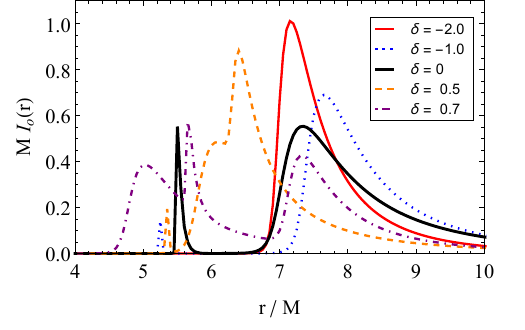}
		\includegraphics[scale=0.68]{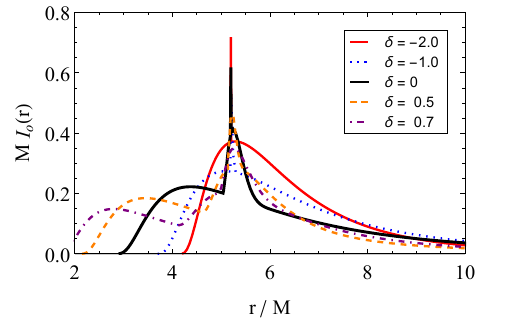}
		\caption{Intensity profiles in the reference frame of the emitter (left panel), and observed intensity profiles for the ISCO disk model (middle panel) and the EH disk model (right panel) for different values of $\bar\delta$.}
		\label{fig:obsprofile}
	\end{figure}

	In the Mathematica-based ray-tracing code used to produce the results that follow, the accretion disk in the equatorial plane is modeled by a monochromatic intensity profile that follows the Gralla-Lupsasca-Marrone (GLM) model \cite{Gralla:2020srx}, which in the reference frame of the emitter takes the form
	\begin{equation}
		I_e\left(r; \gamma, \beta,\sigma\right)=\frac{\exp\left\{-\frac{1}{2}\left[\gamma+\text{arcsinh}\left(\frac{r-\beta}{\sigma}\right)\right]^2\right\}}{\sqrt{\left(r-\beta\right)^2+\sigma^2}},
	\end{equation}
	where $\gamma$, $\beta$, and $\sigma$ are free parameters that control the shape of the emission profile. In this work we consider two different disk models based on the GLM model: the first, which we denote the ISCO model, is motivated by the fact that circular orbits in the spacetimes considered in this manuscript become unstable for orbital radii smaller than the radius of the Innermost Stable Circular Orbit (ISCO), which stands at $r_{ISCO}=6M$, and thus the intensity profile is expected to peak at this radius and rapidly decrease for smaller radii; the second, which we denote the EH model, is motivated by the fact that even if stable circular orbits do not exist in the region between the ISCO and the EH, the infalling of matter towards the EH should contribute to the emitted intensity profile, thus leading to a peak of intensity closer to the EH. The ISCO disk model is characterized by the parameters $\gamma=-2$, $\beta=6M$, and $\sigma=M/4$, whereas the EH disk model is characterized by the parameters $\gamma=-3$, $\beta=2M$, and $\sigma=M/8$. The emitted intensity profiles are plotted in the left panel of Fig.\ref{fig:obsprofile}.

\begin{figure}
		\includegraphics[scale=0.32]{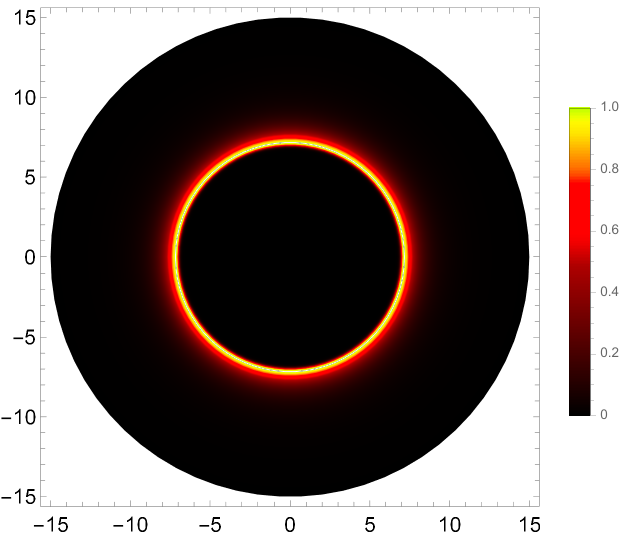}
		\includegraphics[scale=0.32]{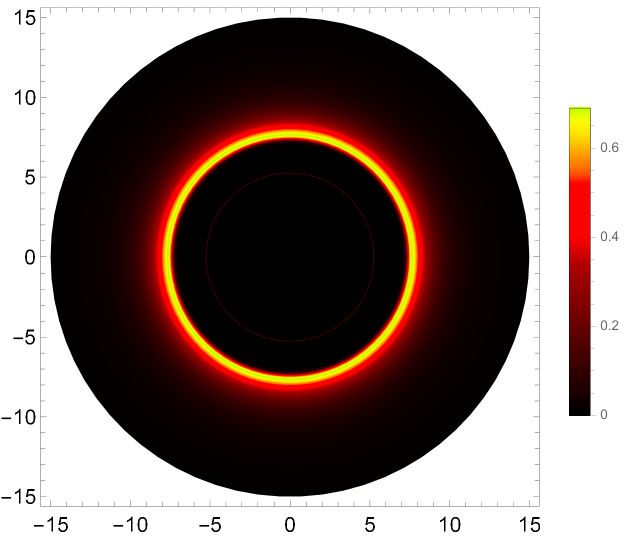}
		\includegraphics[scale=0.32]{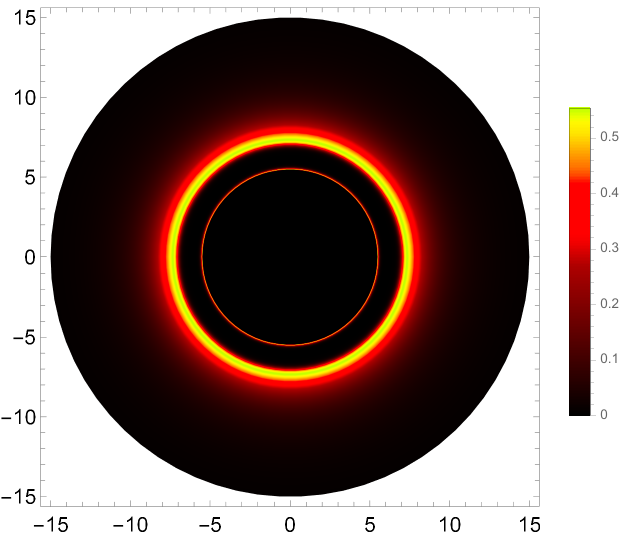}
		\includegraphics[scale=0.32]{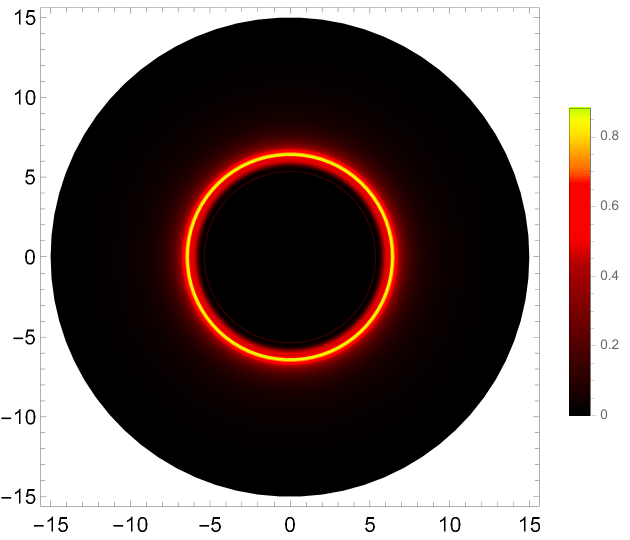}
		\includegraphics[scale=0.32]{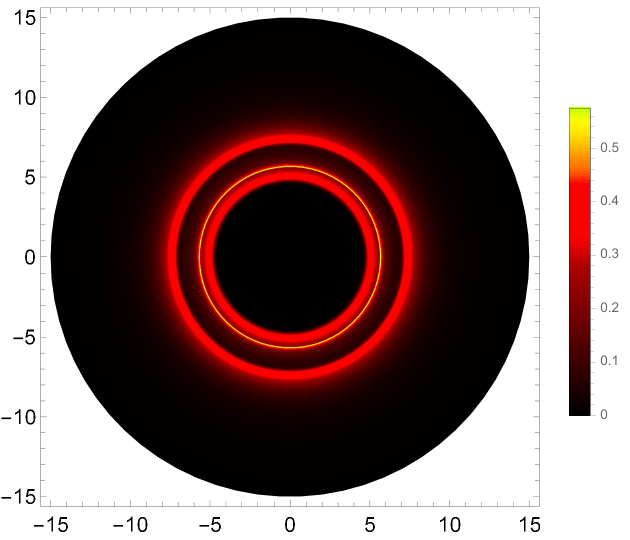}\\
		\includegraphics[scale=0.32]{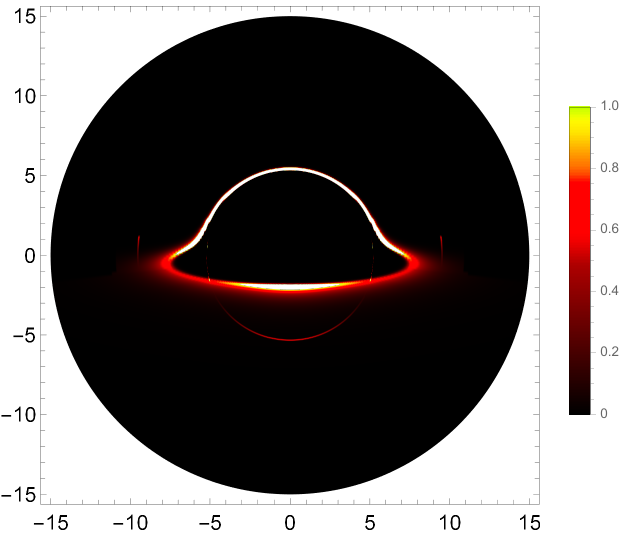}
		\includegraphics[scale=0.32]{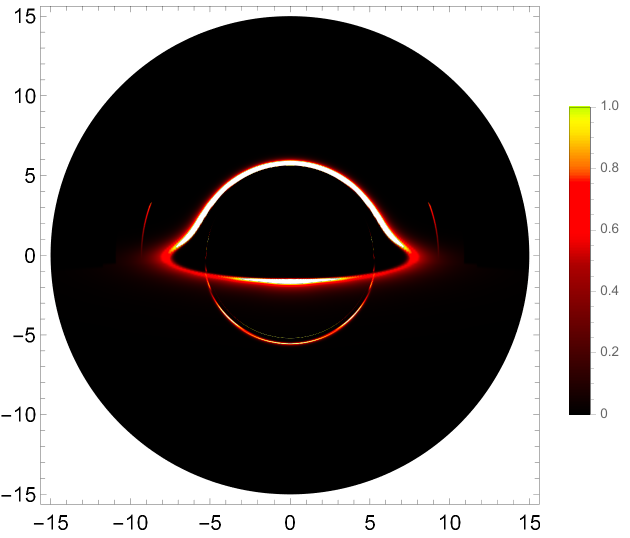}
		\includegraphics[scale=0.32]{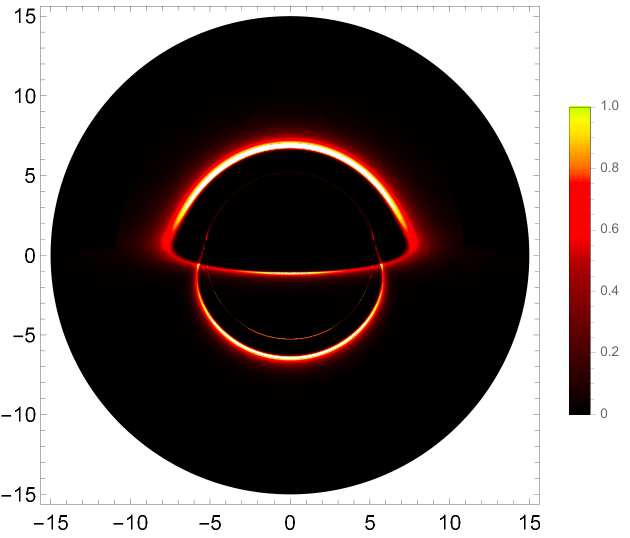}
		\includegraphics[scale=0.32]{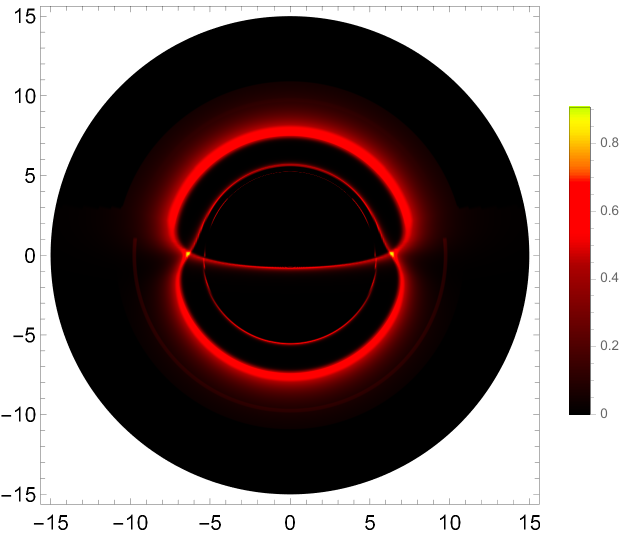}
		\includegraphics[scale=0.32]{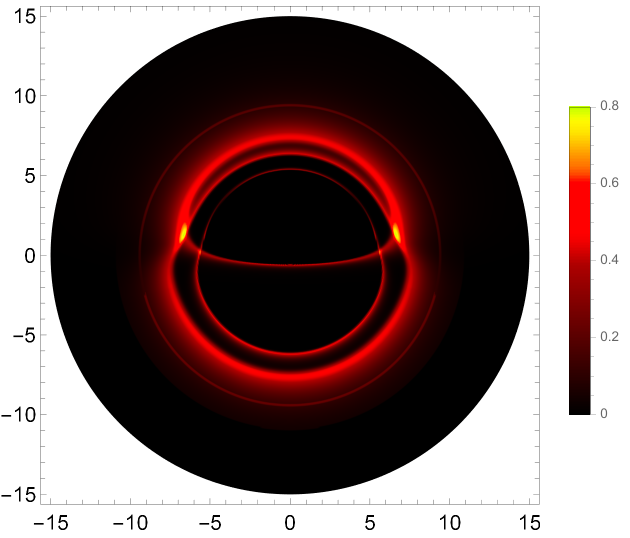}\\
		\includegraphics[scale=0.32]{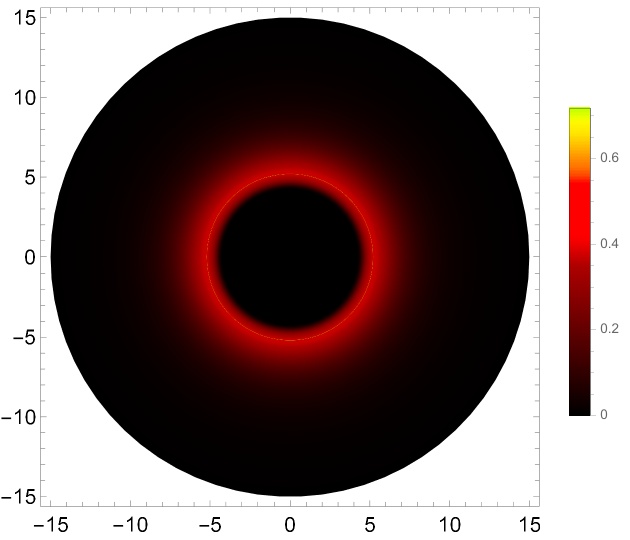}
		\includegraphics[scale=0.32]{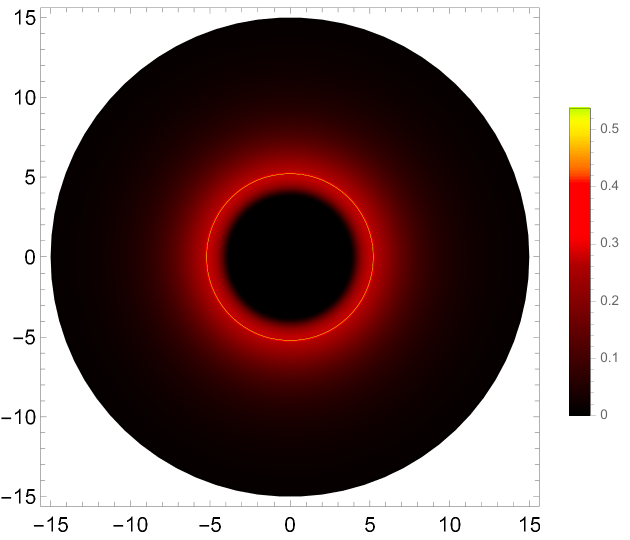}
		\includegraphics[scale=0.32]{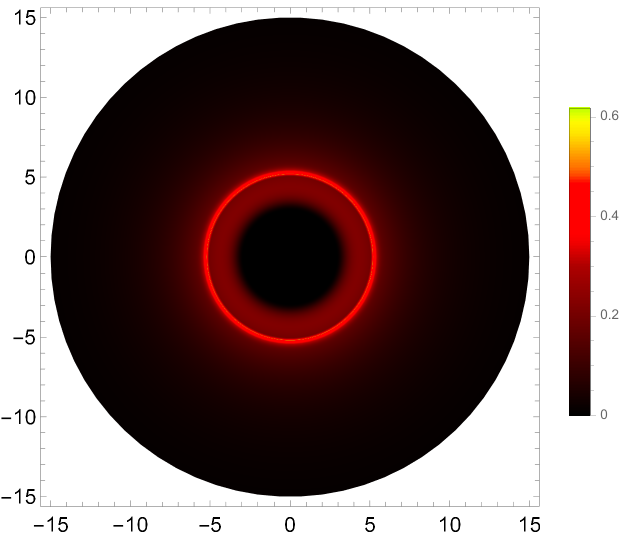}
		\includegraphics[scale=0.32]{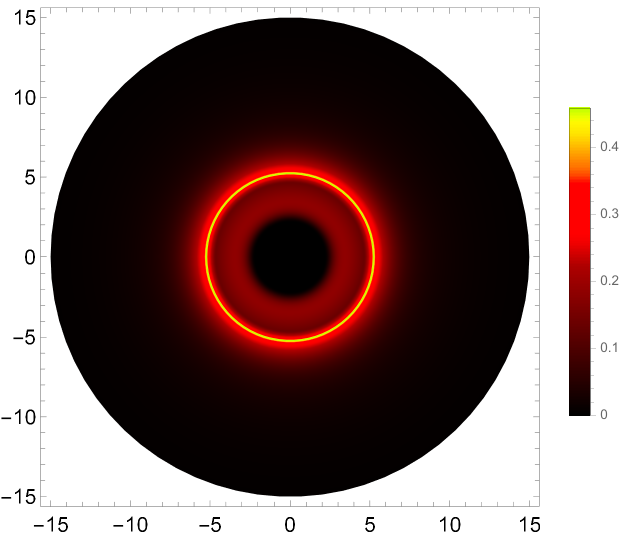}
		\includegraphics[scale=0.32]{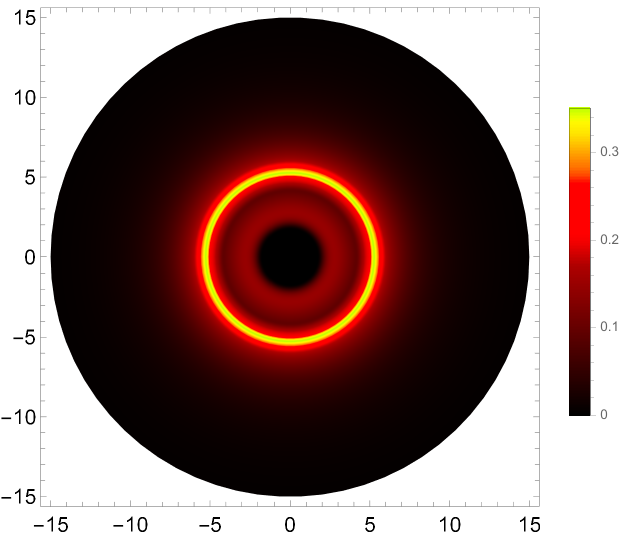}\\
		\includegraphics[scale=0.32]{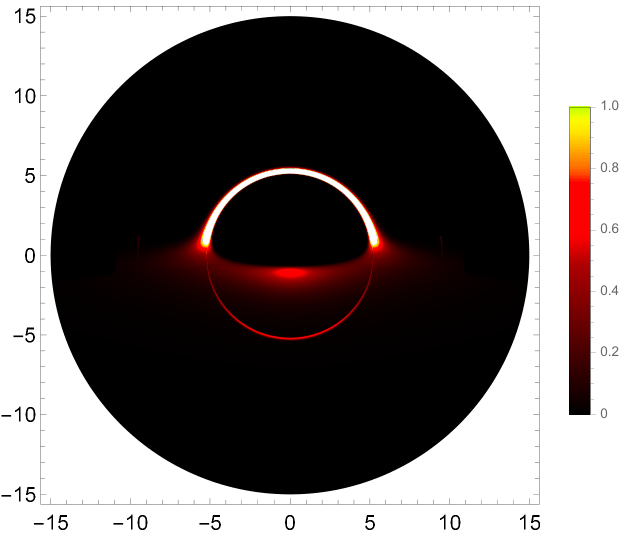}
		\includegraphics[scale=0.32]{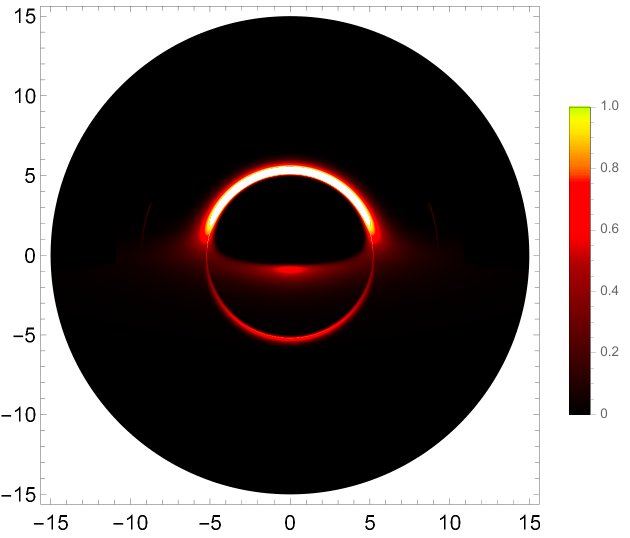}
		\includegraphics[scale=0.32]{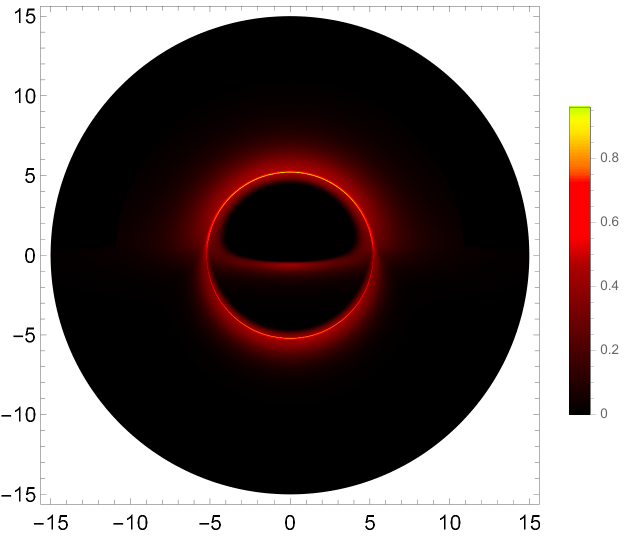}
		\includegraphics[scale=0.32]{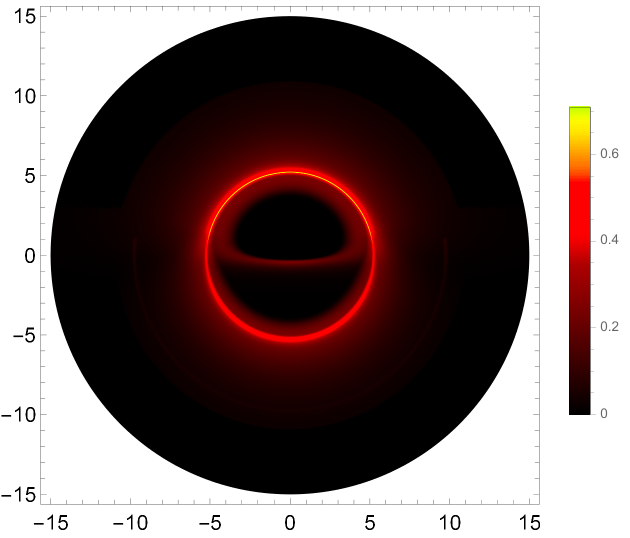}
		\includegraphics[scale=0.32]{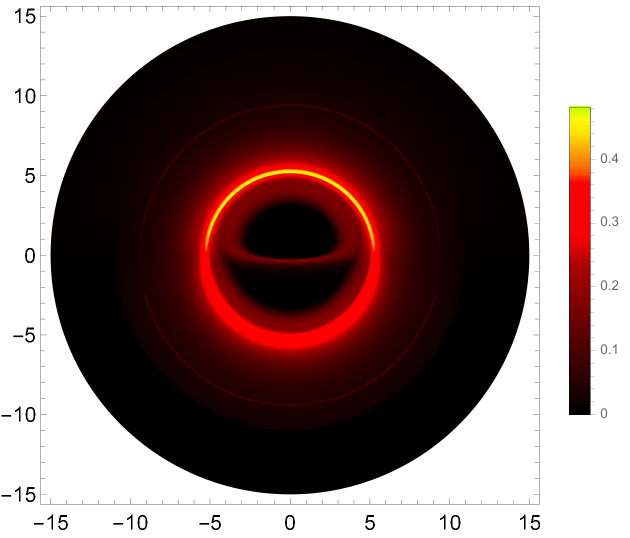}
		\caption{Observed images for the ISCO disk model (top two rows) and the EH disk model (bottom row rows) with inclination angle of $0^\circ$ (top) and $80^\circ$ (bottom), for $\bar\delta=\{-2; -1; 0; 0.5; 0.7\}$, from left to right.}
		\label{fig:shadow}
	\end{figure}

	In the reference frame of the observer, the observed intensity profiles $I_o$ are redshifted with respect to the emitted intensity profiles $I_e$, an effect that takes into consideration the background metric where the photons are propagating. Indeed, the observed intensity profiles are given by
	\begin{equation}
		I_o\left(r\right)=g_{tt}^2 I_e\left(r\right).
	\end{equation}
	The observed intensity profiles for both the ISCO disk model and the EH disk model are given in the middle and right panels of Fig.\ref{fig:obsprofile}, respectively, for different values of the free parameter $\bar\delta$, where the thick black line represents the Schwarzschild solution. The corresponding observed images are shown in Fig.\ref{fig:shadow}. The results indicate that negative values of $\bar\delta$ result in a reduction of the light-ring contribution to the observed profiles and images, while the opposite effect occurs for positive values of $\bar\delta$. This result is consistent with the expectation from the analysis of the geodesic congruences in these spacetimes, as negative values of $\bar\delta$ have a repulsive effect, thus reducing the number of photons that reach the observer after approaching the photon sphere, and vice-versa for positive values of $\bar\delta$. Furthermore, the size of the black-hole shadow is also affected by the parameter $\bar\delta$, with positive values of this parameter resulting in an increase in the size of the shadow and negative values resulting in a decrease of the size of the shadow. For the ISCO model, it can even be observed that for $\bar\delta=0.9$ an additional secondary image appears outside of the light-ring due to the increased light bending.

	\section{Conclusions}\label{sec:Conclusion}
	In this work, we derived the field equations for spherical symmetric spacetimes in 1PNGR by using the most general tetrad for spherical symmetry. These equations were decomposed in their symmetric and antisymmetric parts, and three branches of solutions to the antisymmetric field equations were found.

	We then discussed vacuum solutions for all three branches. The first branch is analyzed in Sec. \ref{sssec:br1}. In this branch, one observes that after a time-coordinate transformation, the metric can be reduced to a Schwarzschild form. The second branch is analyzed in Sec. \ref{sssec:br2}. Similarly, after a time-coordinate transformation, the metric takes the form of a mildly modified Schwarzschild metric, see Eq. \eqref{eq:metric0}, which renders the metric asymptotically non-flat. Since this branch is more difficult to analyze, we presented the most general metric solution for the metric degrees of freedom, but not the most general solution for the full tetrad. Assuming that the tetrad is independent of the time coordinate, an analytic solution for the remaining degree of freedom $\beta=\beta(r)$ could be found, see \eqref{eq:betaII}. We emphasize that the tetrad considered encompasses a particular case used in previous works for $f(T)$ gravity and scalar-tensor theories, where exact and numerical black hole solutions have been found ~\cite{Bahamonde:2021srr,Bahamonde:2022lvh,Bahamonde:2022chq}. Finally, the third branch was analyzed in Sec. \ref{sssec:br3}. This branch is more involved and also quite difficult to solve in general. In order to simplify the equations, we considered the static case with $g_{rr}=-1/g_{tt}$ which corresponds to a Schwarzschild metric and reduces to the Minkowski metric for $M\rightarrow 0$.

	We also studied some phenomenological aspects of the analytic solution we found for the second branch which involves two free parameters $\epsilon$ and $\alpha_{0}$. In particular, we found that there is an event horizon at $r_{H}=2M$, we calculated the Komar mass, which attains a teleparallel correction, and found a curvature singularity at $r=0$. We showed that modification of the classical phenomena of light deflection, Shapiro delay and the perihelion shift can easily be expressed in terms of their value in Schwarzschild geometry. These can be used to find constrains on the teleparallel parameters.	Moreover, we gave a detailed study of the photon sphere and the black hole shadow. We realized that our results are in agreement with the results from the analysis of the congruences of geodesics. In particular, we have shown that, depending on the values chosen for the free parameters of the solution, one might observe additional attractive or repulsive effects on the propagation of photons, which could result in a non-negligible observable distortion of the primary, secondary, and light-ring components of the observed image, and in a modification of the size of the black-hole shadow. In the future, the next generation of long base-line interferometers like the ngEHT could provide observations precise enough to allow one to constrain the values of the free parameters of the models considered in this work with experimental data.

	Moreover, we could derive the perturbed equations at linear or higher order around these background solutions in order to study the perturbations of black holes or other spherically symmetric compact objects in this theory, and then analyze the emission of gravitational waves in the ringdown phase of merger events, which are characterized by the quasinormal mode frequencies. Comparing these to observations in the last stage of the gravitational wave signal may show how these modes compare to their GR counterparts. This analysis would not only provide additional information about the source of the gravitational waves, but also test and constrain the gravitational theories we consider. The propagation of gravitational waves in teleparallel gravity has been already studied \cite{Hohmann:2018jso}.

	\begin{acknowledgments}

		 SB was supported by ``Agencia Nacional de Investigación y Desarrollo" (ANID),  Grant ``Becas Chile postdoctorado al extranjero" No. 74220006.
	MH gratefully acknowledges the full financial support by the Estonian Research Council through the Personal Research Funding project PRG356.
	CP was funded by the cluster of excellence Quantum Frontiers funded by the Deutsche Forschungsgemeinschaft (DFG, German Research Foundation) under Germany's Excellence Strategy - EXC-2123 QuantumFrontiers - 390837967. The authors would like to acknowledge networking support by the COST Action CA18108.
		JLR acknowledges the European Regional Development Fund and the programme Mobilitas Pluss for financial support through Project No.~MOBJD647, and project No.~2021/43/P/ST2/02141 co-funded by the Polish National Science Centre and the European Union Framework Programme for Research and Innovation Horizon 2020 under the Marie Sklodowska-Curie grant agreement No. 94533.

	\end{acknowledgments}

	\bibliographystyle{utphys}
	\bibliography{references}

\end{document}